\documentclass[usenatbib,usegraphicx]{mn2e}

\usepackage{subfigure}
\usepackage{longtable}
\usepackage{graphicx}
\usepackage{graphics}
\usepackage{keyval}
\usepackage{trig}
\usepackage[american]{babel}
\usepackage{url}
\usepackage{color}
\usepackage{amsmath}
\usepackage{bm}
\usepackage{longtable}

\def\beq{\begin{equation}}
\def\eeq{\end{equation}}
\def\bey{\begin{eqnarray}}
\def\eey{\end{eqnarray}}

\def\msun{M_\odot}

\def\lsim{\mathrel{\raise.3ex\hbox{$<$\kern-.75em\lower1ex\hbox{$\sim$}}}}
\def\gsim{\mathrel{\raise.3ex\hbox{$  $\kern-.75em\lower1ex\hbox{$\sim$}}}}

\def\tz{t_\mathrm{0}}
\def\te{t_\mathrm{E}}
\def\tE{t_\mathrm{E}}
\def\rhostar{\rho_\star}
\def\uz{u_\mathrm{0}}

\def\thetae{\theta_\mathrm{E}}
\def\fs{F_\mathrm{S}}
\def\fb{F_\mathrm{B}}
\def\ds{D_\mathrm{S}}
\def\dl{D_\mathrm{L}}

\def\a0{A_{\mathrm{0}}}

\def\sin{s_\mathrm{in}}
\def\sout{s_\mathrm{out}}
\def\tin{t_\mathrm{in}}
\def\tout{t_\mathrm{out}}

\def\dtcc{{\Delta t_{\rm cc}}}

\newcommand{\bs}{\mbox{$\!\!\!$}}
\newcommand\prior{\pi}
\newcommand\priorp[1]{\prior\left(#1\right)}
\newcommand\phit{\prior_{\rm H}}

\newcommand\Eq[1]{Eq.~(\ref{#1})}
\newcommand\Fig[1]{Fig.~\ref{#1}}
\newcommand\Tab[1]{Table~\ref{#1}}

\title{A Bayesian algorithm for model selection
	applied to caustic-crossing binary-lens microlensing events}
\author[]
{	N. Kains$^{1}$ \thanks{email:nkains@eso.org},
	P. Browne$^{2}$,
	K. Horne$^{2}$,
	M. Hundertmark$^{2}$,
	A. Cassan$^{3}$\\
\\
$^{1}$European Southern Observatory,
	Karl-Schwarzschild Stra\ss e 2,
	85748 Garching bei M\"{u}nchen, Germany
\\
$^{2}$SUPA, School of Physics and Astronomy,
	University of St. Andrews, North Haugh,
	St Andrews, KY16 9SS, United Kingdom
\\
$^{3}$Institut d'Astrophysique de Paris,
	UMR7095 CNRS-Universit\'{e} Pierre \& Marie Curie,
	98 bis boulevard Arago,
	75014 Paris, France
}

\begin{document}

\date{Accepted ... Received ... ; in original form ...}

\pagerange{\pageref{firstpage}--\pageref{lastpage}} \pubyear{2012}

\maketitle

\label{firstpage}

\begin{abstract} We present a full Bayesian algorithm designed to perform 
automated searches of the parameter space of caustic-crossing binary-lens 
microlensing events. This builds on previous work implementing priors derived 
from Galactic models and geometrical considerations. The geometrical 
structure of the priors divides the parameter space into well-defined boxes that 
we explore with multiple Monte Carlo Markov Chains. We outline our Bayesian 
framework and test our automated search scheme using two data sets: a synthetic 
lightcurve, and the observations of OGLE-2007-BLG-472 that we analysed in 
previous work. For the synthetic data, we recover the input parameters. For 
OGLE-2007-BLG-472 we find that while $\chi^2$ is minimised for a planetary mass-ratio
model with extremely long timescale, the introduction of priors and minimisation 
of BIC, rather than $\chi^2$, favours a more plausible lens model, a binary star 
with components of $0.78$ and 0.11~$\msun$ at a distance of $6.3$ kpc, compared 
to our previous result of $1.50$ and $0.12~\msun$ at a distance of 1~kpc.

\end{abstract}


% =====================================================
\section{Introduction}\label{sec:intro}
% =====================================================

Gravitational microlensing \citep{einstein36} is a well-established technique to 
detect extrasolar planets (e.g. \citealt{maopaczynski91}, \citealt{beaulieu06}, 
\citealt{muraki11}), and is complementary to other methods, being able to 
probe low-mass cool planets that are inaccessible to them from the ground. This allows us to carry out statistical studies of planets of all masses located at a few AU from their host star \citep{cassan12}.
Microlensing occurs when one or several compact objects are located between a 
source star and the observer, leading to a gravitational deflection of the light 
from the source star by the ``lens" objects. As the source and lens move in and 
out of alignment, this deflection is observable in the form of a simple 
characteristic brightening and fading pattern when the lensing object is a single 
star (\citealt{paczynski86}), but takes a much more complex form when the lens is 
made up of more than one object. When that happens, the lightcurve typically 
features ``anomalies", which can be modelled to determine the nature of the 
lensing system. One of the configurations that can lead to anomalies is when the 
lensing system contains one or more planets. In order to determine the properties 
of these planets, the anomalies must be analysed through detailed modelling; this 
paper is concerned with cases where the lens consists of two components.

Analysing anomalous microlensing lightcurves can be a significant computational 
challenge for a number of reasons. The calculation of a full binary-lens 
lightcurve, including the effects of having an extended source, is an expensive 
process computationally, and the parameter space to be explored is complex, with 
several degeneracies (e.g. \citealt{kubas05}). This is the case even when 
second-order effects, such as that of parallax due to the Earth's orbit or 
orbital motion in the lensing system, are ignored.

A significant number of the $\sim 1500$ microlensing events 
now being discovered by survey teams in a season exhibit 
anomalies due to stellar or planetary companions to the lens star.
Many of these are caustic-crossing events in which
the lightcurve exhibits rapid jumps, brightening when a new pair of images forms 
and fading when two images merge and disappear. 
\cite{cassan08} introduced an advantageous parameterisation for
caustic-crossing events by linking two 
parameters, $\tin$ and $\tout$, to the caustic-crossing times and
two parameters, $\sin$ and $\sout$, to the ingress and 
egress points where the source-lens trajectory crosses the caustic curve.
These parameters make it easier to locate all possible source-lens
trajectories that fit the observed caustic-crossing features.

\cite{kains09} used the \cite{cassan08} parameters to analyse
the observed lightcurve of the microlensing event OGLE-2007-BLG-472,
which exhibits two strong caustic-crossing features separated by about 3 days.
This short duration suggested that the anomaly could be due to the 
source crossing a small planetary caustic, motivating detailed modelling 
to rule out alternative binary star lens models.
The lowest-$\chi^2$ model has a planetary mass ratio, but
an extremely long event timescale, $\te \sim 2000$ days, 
much longer than the 2-200~day range typical of Galactic 
Bulge microlensing events. On this basis \cite{kains09} 
rejected the global $\chi^2$ minimum by placing an ad-hoc 300~day cutoff on 
$\te$, and suggested that a Bayesian approach including 
appropriate priors on all the parameters would more naturally shift
the posterior probability to local $\chi^2$ minima with less extreme parameters.

\cite{cassan09} derived analytic formulae for the prior $\priorp{\sin,\sout}$ 
corresponding to a uniform and isotropic distribution of lens-source 
trajectories, which are specified by an angle $\alpha$
and impact parameter $\uz$.
A suitable prior on $\te$ arises by using a model of microlensing in the 
Galaxy to determine distributions for the lens and source distances
and their relative proper motion, or alternatively by using a parameterised
model fitted to the observed distribution of $\te$ among all the events found in 
the microlensing survey. In either case a prior on $\te$ effectively penalises 
very long and very short events, lowering the posterior probability
of the $\te\sim2000$~d global $\chi^2$ minimum found for 
OGLE-2007-BLG-472 and favouring local minima with more typical event 
timescales. Priors on other parameters can also be derived from models of stellar population synthesis such as the Besan{\c c}on model \citep{robin03}, which we use in this work.

In this paper we develop further the Bayesian analysis of caustic-crossing 
events, exploiting intrinsic features of the $\priorp{\sin,\sout}$ prior to 
specify and 
test a procedure suitable for automatic exploration 
of the full parameter space. We test the procedure using synthetic lightcurve 
data, and we re-analyse the OGLE-2007-BLG-472 data to compare the results of maximum 
likelihood analysis ($\chi^2$ minimisation) with the full Bayesian analysis 
including appropriate priors.

\section{Binary-lens microlensing}

 In the context of microlensing, caustics are locations in the source plane, 
behind the lens, where the magnification is infinite. A point-mass lens
produces a point caustic directly behind the lens where formation
of an Einstein Ring gives infinite magnification for a point source,
or very large magnification for a finite size source. 
The point-lens gives a symmetric magnification pattern $A(u)$, with $u$ the 
projected source-lens distance in the source plane, in units of the Einstein Ring radius.
The linear source trajectory has impact parameter $\uz$ and a timescale $\te$, 
both expressed in units of the angular Einstein radius \citep{einstein36},

\begin{equation}
\label{eq:thetae}
	\thetae = \sqrt{ \frac{ 4\,G\,M }{ c^2 }
	\left( \frac{ D_\mathrm{S} - D_\mathrm{L} }
	{D_\mathrm{S}\, D_\mathrm{L} } \right)}
\ ,
\end{equation}

\noindent
where $M$ is the lens mass, and $\ds$ and $\dl$ are the distances to the source 
and the lens respectively.
This produces a symmetric lightcurve with magnification 
$A(u(t))$ peaking at $A_0$ at time $\tz$. Thus 3 parameters, $\uz$, $\tz$, and 
$\te$ define the shape of a point-source point-lens lightcurve.
Finite source effects alter the peak of the lightcurve
when $\uz$ is of order $\rhostar=\theta_\star/\thetae$,
the source star radius in Einstein radius units.

 With two or more lens masses, the simple point caustic becomes a more complex 
set of closed curves consisting of concave segments joining at cusps, the shapes 
and locations depending on the lens masses and locations. Microlensing lightcurve 
anomalies, relative to the point-lens model, arise from the asymmetric 
magnification pattern associated with these caustic curves. The source trajectory 
may pass nearby to a cusp, causing a bump in the lightcurve, or cross over a 
caustic curve, resulting in a variety of complex lightcurve features, depending 
on the exact lens geometry. For a static binary lens, the caustic pattern 
depends on the mass ratio $q$ and separation $d$ between the two lens masses.
The source trajectory relative to the caustics is specified by the impact 
parameter $\uz$ relative to the centre of mass of the lens, and the trajectory 
angle $\alpha$ relative to the line connecting the two lens masses.

 As emphasised by \cite{cassan08} and \cite{kains09},
for caustic-crossing events the standard parameterisation makes it very 
difficult to conduct a systematic exploration of the parameter space. 
The alternative parameterisation formalised by 
\cite{cassan08} replaces ($\uz$, $\alpha$, $\tz$, $\te$, $\rhostar$)
by equivalent parameters ($\sin$, $\sout$, $\tin$, $\tout$, $\dtcc$) 
that are more closely related to observable 
lightcurve features, and therefore better constrained by observations.
Of the ``standard" binary-lens microlensing parameters, two are retained: the 
mass ratio of the lens components $q$ ($\leq 1$), and their projected separation 
$d$.

 \cite{kains09} show that the alternative parameters are better suited to 
fitting caustic-crossing event lightcurves, finding models that are widely 
separated and easily missed with the standard binary-lens parameterisation. 
However, the global $\chi^2$ minimum found in the \cite{kains09} analysis of 
OGLE-2007-BLG-472 was a model with a ``planetary" mass ratio $q\sim10^{-4}$, 
but with an extremely long timescale, $\te\sim2000$ days. This model 
was rejected through a qualitative discussion with the expectation that
a Bayesian analysis would more naturally shift the best fit to 
a different local $\chi^2$ minimum with less exotic parameters.
In this paper, we add priors on relevant parameters, attempting to remove the 
need for such qualitative arguments by using a badness-of-fit statistic that 
includes both the likelihood and additional terms originating from prior 
information.

\section{Data}

\subsection{Synthetic Lightcurve Data}

\begin{figure}
  \centering
  \includegraphics[width=6cm, angle=270]{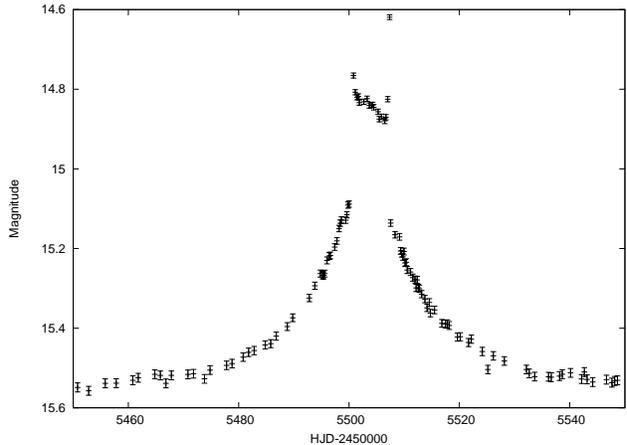}
  \caption{Synthetic data used in this paper, plotted with 1-$\sigma$ 
 error bars calculated using \Eq{eq:sigma}.
 \label{fig:data}}
\end{figure}

To test our automated algorithm, we generated a data set using the parameters 
given in \Tab{tab:888_trueparameters}, selected to reproduce features seen in observed 
anomalous microlensing lightcurves. The chosen parameters correspond to a 
caustic-crossing binary-lens event with crossings separated by 7 days and 
occurring near the lightcurve peak (Fig.~\ref{fig:data}).

% =====================================================
\begin{table}
\begin{center}
  \begin{tabular}{ccc}
\hline
    Parameter & & Units
\\ \hline
   $\tz$ & $5503.6$ & MHJD
\\ $\te$ & $27.2$ & days
\\ $\alpha$ & $1.68$ & rad
\\ $\uz$ & $0.1$ & $-$
\\ $\rhostar$ & $0.003$ & $-$
\\ $d$  & $1.22$ & $-$
\\ $q$ 	& $0.08$ & $-$
\\ $g=\fb/\fs$ & $5$ & $-$
\\ \hline
  \end{tabular}
  \caption{Standard binary-lens parameters used to generate our synthetic data. 
\label{tab:888_trueparameters}}
  \end{center}
\end{table}
% =====================================================

For an observation at time $t_i$, when the source is magnified by a factor 
$A(t_i)$, the true model magnitude is

\begin{equation}
\label{eq:mi}
	\mu_i = -2.5\, \mathrm{log_{10}}( \fs\, A(t_i) + \fb )
\ ,
\end{equation}

\noindent
where the un-magnified source flux $\fs$ was chosen to be 1/5 of the blend flux 
$\fb$, which represents un-magnified stars that are blended with the microlensing 
target. The source-lens trajectory's impact parameter $\uz=0.1$ is small enough 
to reach magnification $A\sim10$ near the closest approach at time $\tz$.

We obtain synthetic magnitude data $m_i$ by using a pseudo-random number 
generator to sample a Gaussian distribution with mean $\mu_i$ and standard 
deviation $\sigma_i$, given by

\begin{equation}
\label{eq:sigma}
	\sigma_i = \frac{0.01}{1+\left| m_0 - \mu_i \right|}
\ ,
\end{equation}

\noindent 
where $m_0=-2.5\, \mathrm{log_{10}}(\fs+\fb)$ is the baseline magnitude, 
corresponding to the un-magnified source flux plus the blend flux. The fractional 
error bars are thus 1\% at the baseline and decrease when the source is 
magnified. After generating the synthetic magnitude data, we re-scaled these 
error bars to obtain a $\chi^2$ of 1 per degree of freedom for the true model. 
This approximates the common practice of rescaling the nominal error bars when 
fitting to observed microlensing lightcurves.

We employed a non-uniform cadence emulating a typical microlens observing 
strategy. We start with a baseline cadence of one observation per night, 
increasing to 3 observations per night as the event nears the peak predicted by a 
point-source point-lens (PSPL) fit to the earlier data. When the anomaly is 
detected, i.e. when the synthetic lightcurve data departs significantly from the 
PSPL fit, the cadence increases to 5 observations per night. From the resulting 
lightcurve, a random sample of $N$ points is selected to emulate data loses, e.g. 
due to bad weather or technical issues.

\begin{figure}
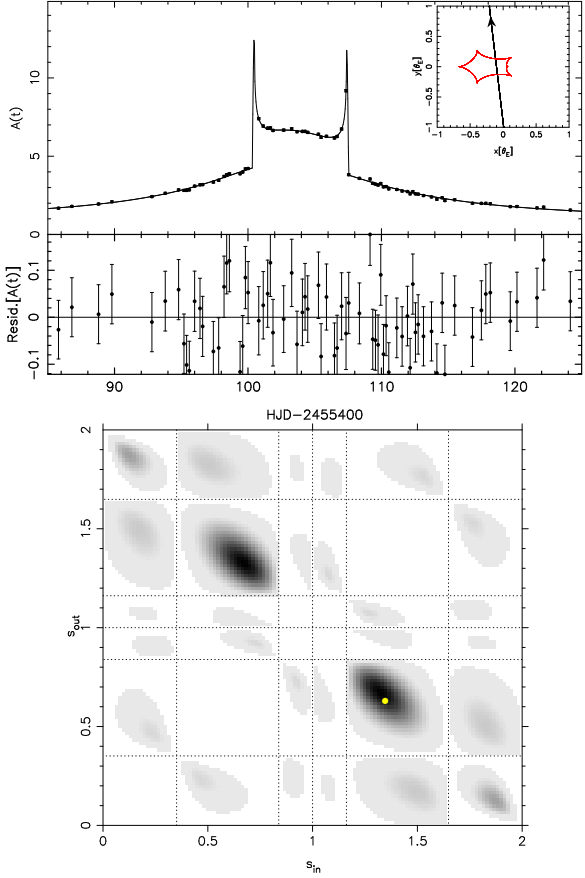

  \centering
  \includegraphics[width=6cm, angle=270]{fig/OB08888-model0-plotblfit.ps}
 \hspace{1cm} \includegraphics[width=6cm, angle=270]{fig/OB08888-model0-pmap.ps}
  \caption{\textit{Top}: Synthetic data and true model lightcurve, 
generated with the parameters given in \Tab{tab:888_trueparameters}. 
\textit{Bottom}: The ($\sin, \sout$) prior map, with the location of the 
true model shown with a filled yellow circle. 
\label{fig:888_truelc}}
\end{figure}

The resulting synthetic lightcurve, retaining 199 data points, is 
shown on \Fig{fig:data}.   A plot of the true model lightcurve with the 
parameters given in \Tab{tab:888_trueparameters} is shown on \Fig{fig:888_truelc}.

% =====================================================
% =====================================================

% =====================================================
% =====================================================
 
 \subsection{OGLE-2007-BLG-472}
 
This event was alerted during the 2007 microlensing observing season by the OGLE 
collaboration, and followed up from two observing sites by the PLANET 
collaboration. It was used as the test event by Kains et al. (2009, see that 
paper for full details on the data sets) to illustrate the capabilities of a 
modelling scheme based on the parameters defined by \cite{cassan08}. We use the 
same event here for comparison and in particular to show that the full Bayesian 
analysis shifts the posterior probability away from the exotic parameters found 
in the previous maximum likelihood analysis.

\section{Bayesian framework}

Our analysis implements a Bayesian framework for fitting microlensing 
events involving caustic crossings. For $M$ parameters $\theta$ and data $D$, the 
posterior probability density in the $M$-dimensional parameter space is
\begin{equation}
	P(\theta|D) = \frac{
	P(D|\theta)\, \priorp{\theta}
	}{ \int P(D|\theta)\, \priorp{\theta}\, d^M \theta}
\ .
\end{equation}

\noindent
Here $\priorp{\theta}$ is the prior on the $M$ parameters
and $P(D|\theta)$ is the likelihood, e.g.\,
for Gaussian errors with known standard deviations $\sigma_i$
the likelihood is 
\begin{equation}
	L(\theta) \propto P(D|\theta) = \frac{
	\exp{ \left\{ -\frac{1}{2}\chi^2(\theta,D) \right\} }
	}{Z_D}
\ ,
\end{equation}
with
\begin{equation}
\chi^2(\theta,D)=\sum_{i=1}^N 
	\left( \frac{ D_i-\mu_i(\theta)}{\sigma_i} \right)^2
\ ,
\end{equation}
where $\mu_i(\theta)$ is the model prediction for data $D_i$,
and
\begin{equation}
	Z_D = \left( 2\,\pi \right)^{N/2}\, \prod_{i=1}^N \sigma_i
\ ,
\end{equation}
\noindent is a measure of the $N$-dimensional volume admitted by the data.

In fitting the binary lens model to microlensing lightcurve data,
we project the posterior distribution in the full 
$M$-dimensional parameter space onto the $(d,q)$ plane, 
a process known as {\it marginalising} over the 
$m= M-2$ {\it nuisance} parameters, which we denote 
collectively by $\beta$:
\begin{equation}\label{eq:subset}
\begin{array}{rl}
	P(d,q|D) & \bs = \int P(d,q, \beta|D)\, d^m\beta
\\ \\	& \bs = \priorp{d,q}\, \int P(\beta |D,d,q)\, d^m\beta
\ ,
\end{array}
\end{equation}
where $\priorp{d,q}$ is the prior distribution on the $(d,q)$ plane.
We take $\priorp{d,q}$ to be uniform in log ${d}$ and log ${q}$. This choice comes from the fact that the sizes of the caustics behave like power laws of $d$ and $q$.

We then marginalise over
nuisance parameters by simply averaging over the samples of our Markov Chain Monte Carlo algorithm (MCMC, see e.g.\citealt{gelmanbook} for more background information on this),
\begin{equation}
	\int  X(\theta)\, P(\beta|D,d,q)\, d^m\beta
	\approx \left< X \right>
\ ,
\end{equation}
where $X(\theta)$ is any function of the parameters, and 
we use the notation $\left< X \right>$ to refer to
a simple unweighted average over the MCMC samples. 
The result is a map of the posterior probability distribution $P(d,q|D)$.
We will find that the maximum aposteriori (MAP) estimates of $(d,q)$,
which maximise $P(d,q|D)$,
can be quite different from the maximum likelihood (ML) estimates, 
which maximise $P(D|q,d,\beta)$, or mimimise $\chi^2$.

\subsection{Feature-based Parameters
and Structure of the Prior $\priorp{\sin,\sout}$}

The benefit of using the \cite{cassan08} parameters
$(\sin,\sout,\tin,\tout,\dtcc)$, rather than the standard parameters
$(\uz,\alpha,\te,\tz,\rhostar)$, is two-fold.
First, the caustic-crossing time parameters $\tin$, $\tout$ and $\dtcc$ can often 
be tightly constrained by features in the observed lightcurve.
Second, the $(\sin,\sout)$ parameters bring together onto a
compact square all models that have caustic crossings at those times. In 
contrast, with the standard $(\uz,\alpha,\te,\tz,\rhostar)$ parameters,
the models with caustic crossings at times $\tin$ 
and $\tout$ are widely separated and difficult to locate.

The \cite{kains09} analysis used a genetic algorithm
and assumed uniform priors on the \cite{cassan08} parameters.
This has obvious problems: for example, since the caustic folds 
that make up a caustic structure are concave (see e.g. upper panel inset of 
\Fig{fig:888_truelc}), a linear source trajectory cannot enter and then exit a 
caustic along the same caustic fold. This needs to be reflected in suitable 
priors on the corresponding parameters, in this example, on the 
$(\sin,\sout)$ parameters, which determine where the source-lens trajectory
crosses the caustic folds.

\cite{cassan09} derived analytic formulae for the prior $\priorp{\sin,\sout | \tin, \tout, \dtcc}$, hereafter shortened to $\prior(\sin, \sout)$, 
corresponding to a uniform isotropic distribution of source-lens trajectories, 
and introduced also a prior $\priorp{\te}$ on the event timescale, showing
how $\priorp{\te}$ effectively modifies $\priorp{\sin,\sout}$.
The analytic prior is proportional to the Jacobian of the transformation between 
the standard and \cite{cassan08} parameters,
\begin{equation}
\label{eq:jacobian}
   J =  \left|  \frac{\partial\left(\uz,\alpha,\tE,\tz, \rhostar \right)}
	{\partial\left(\sin,\sout,\tin,\tout,\dtcc \right)} \right|
\ .
\end{equation}
\noindent
\cite{cassan09} evaluated this Jacobian to find the
analytic form of $\priorp{\sin,\sout}$ corresponding to
uniform priors on all standard parameters.

As can be seen in e.g. \Fig{fig:888_truelc}, the prior $\priorp{\sin,\sout}$
covers a compact square, since $\sin$ and $\sout$ run
over the same range as we move around the closed caustic curve.
The square naturally sub-divides into ``sub-boxes", the boundaries of 
which correspond to the cusps. For a caustic 
with $N_c$ cusps, there are thus $N_c^2$ sub-boxes. However, 
the sub-boxes on the anti-diagonal of the $(\sin,\sout)$ square
have $\sin$ and $\sout$ on the same caustic fold,
which cannot occur due to the concave geometry of the folds.
This means that the anti-diagonal sub-boxes must have zero probability.
There are thus $N_c\,(N_c-1)$ sub-boxes to consider for each caustic.

The event timescale prior $\priorp{\te}$
can in principle be obtained by considering models of microlensing in the Galaxy,
mapping the joint distribution of lens mass, lens and source distances,
and relative proper motion onto the corresponding distribution of $\te$
(e.g.\, \citealt{dominik06}).
A convenient alternative is to use the observed distribution of $\te$,
e.g.\, from the OGLE survey.
Caution is needed because of possible biases in fitting $\te$ to 
observed lightcurves, and selection effects lowering the occurrence
of short and long $\te$ events in the survey.
\cite{cassan09} considered two different priors on $\te$
to illustrate their effect on $\priorp{\sin,\sout}$. One was a
distribution of event timescales observed in past microlensing seasons, and 
another was the model distribution of \cite{woodmao95}. These two distributions 
were shown to be in excellent agreement with each other.

In this paper we derive a 2-dimensional joint prior on the event timescale and source size using simulations of synthetic stellar population obtained with the Besan{\c c}on model \citep{robin03}. We briefly describe our method to derive this in the next section. 

\subsubsection{Deriving priors from a Galactic model}

The initial maximum likelihood analysis of the microlensing event OGLE-2007-BLG-472 \citep{kains09} suggests an unusual parameter combination as best description of the data. The strength and plausibility of such an approach can be tested by using a Galactic model that reflects our prior knowledge of the Galactic structure. For interpreting microlensing lightcurves, different Galactic models (\citealt{hangould95b}, \citealt{bennettrhie02}, \citealt{dominik06}) are in use. These models differ in details of the assumed spatial, kinematic, and mass distributions of the Galactic Bulge and Disk stellar populations. A different model which is adapted to reflect the observed star counts in the optical and near-infrared is the so-called  Besan\c{c}on model \citep{robin03}. As indicated by \cite{kerins09}, this model can be used to predict the optical depth of gravitational microlensing events. Moreover it can be used to used to set detailed parameter constraints when combined with the adapted parameter estimates and the source star properties.

Based on the available online catalogue simulation, we generated a sample of stars between Earth and 11 kpc. To ensure that potential lenses, which are typically faint, are included, the apparent magnitude was not constrained. Based on the value used in the previous paper \citep{kains09}, we assumed a visual extinction $A_V=0.7$~mag~kpc$^{-1}$, where the resulting model extinction curve stops increasing after several kpc. For a more accurate description, the calibrated spatial extinction in $K_S$ \citep{marshall06} could have been used, but this would have required a calibration for the  $I$ band, which is typically used in microlensing observations. 

In order to infer microlensing distributions from the simulation, lens-source pairs were randomly drawn from the sample. These were then accepted or rejected depending on the area of their corresponding angular Einstein ring, which gives the instantaneous lensing probability. The simulated bolometric magnitude and effective temperature allowed us to estimate $\rho_*$. Including the simulated proper motion provided us with an estimate for the Einstein time - the only observable parameter directly connected to the lens mass. We did not include the lens-source relative proper motion in the resampling procedure, but this would lead to a distribution that favours shorter Einstein times, as fast lenses lead to larger detection zones on the sky. This is a consequence of moving the lensing cross-section of the instantaneous lensing probability along the lens-source relative proper motion. The correction depends on the annual survey observation and the survey efficiency in tE. For events much longer than the sampling rate, the increase in lensing probability can be modelled  as a stripe on the sky. For a coverage of 240 days which we assume here, the actual prior distribution of the event duration changes its expected value by an amount that is negligible in comparison to the error ellipse of instantaneous case. 

\begin{figure}
  \centering
   \includegraphics[width=8cm, angle=0]{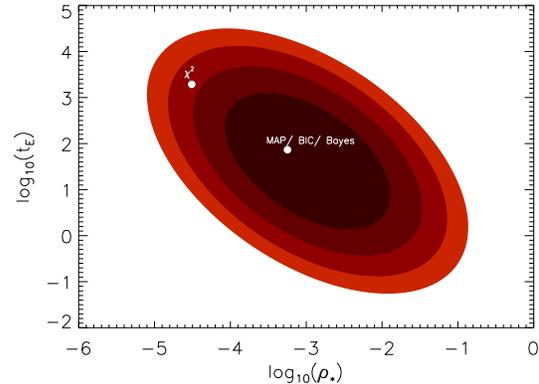}

  \caption{Contour plot of the the logarithm of our joint prior on $\rho_*$ and $\te$. The location of the best-fit models identified for event OGLE-2007-BLG-472 are shown with white filled circles and labelled for the different statistics we use, as discussed in the text. \label{fig:priorcontour}}

\end{figure}

Our estimates illustrate that the value found by \cite{kains09} for $\te$ is much larger and that of $\rho_*$ much smaller than typical samples drawn from the Besan\c{c}on model. Consequently, we determine a bivariate Gaussian prior based on the covariance matrix of $\te$ and angular source star radius of the simulated sample. This joint prior is plotted on \Fig{fig:priorcontour}.

% =====================================================
% =====================================================

Since trajectories requiring very large values of $\te$ and/ or very small $\rho_*$ are suppressed by this 
prior, so too are the corresponding regions of the $(\sin, \sout)$ plane. That 
is, for given 
values of $\tin$ and $\tout$, regions of $(\sin, \sout)$ where the source enters 
and exits the caustic structure very close to the same cusp are suppressed. This 
is evident on the $\priorp{\sin,\sout}$ maps, e.g.\, the bottom panel of 
\Fig{fig:888_truelc},
where the prior is low in the corners along the anti-diagonal line.

Because the \cite{cassan08} parameters assume that the source trajectory 
crosses a caustic, and because we are comparing caustics of different sizes as we 
move across the $(d,q)$ plane,  a full implementation of the uniform isotropic 
prior on source-lens trajectories must account for large caustics being easier to 
hit than small ones. If two models have equal $\chi^2$ but cross caustics of different 
sizes, the prior should favour the model with a larger probability of being hit.
As each $(\sin,\sout)$ corresponds to a different source trajectory angle 
$\alpha$, we quantify this by defining $\phit(d,q,\alpha)$, the 
probability 
that a caustic will be ``hit" by a trajectory with angle $\alpha$. This is 
proportional to the range of impact parameters intersecting the caustic, 
i.e.\, the projected size of the caustic perpendicular to 
the source trajectory. The concave structure of the caustic means that once the 
$N_c$ cusp positions are found, and rotated by an angle $-\alpha$, the vertical 
range then gives the projected cross-section of the caustic.
Thus if the \cite{cassan09} prior $\priorp{\sin,\sout}$ is normalised to 1 when 
integrated over the $(\sin,\sout)$ square, the full prior
multiplies this by $\phit(d,q,\alpha)$.

\subsection{Automated modelling scheme}

\begin{figure*}
  \centering
   \includegraphics[width=11cm, angle=90]{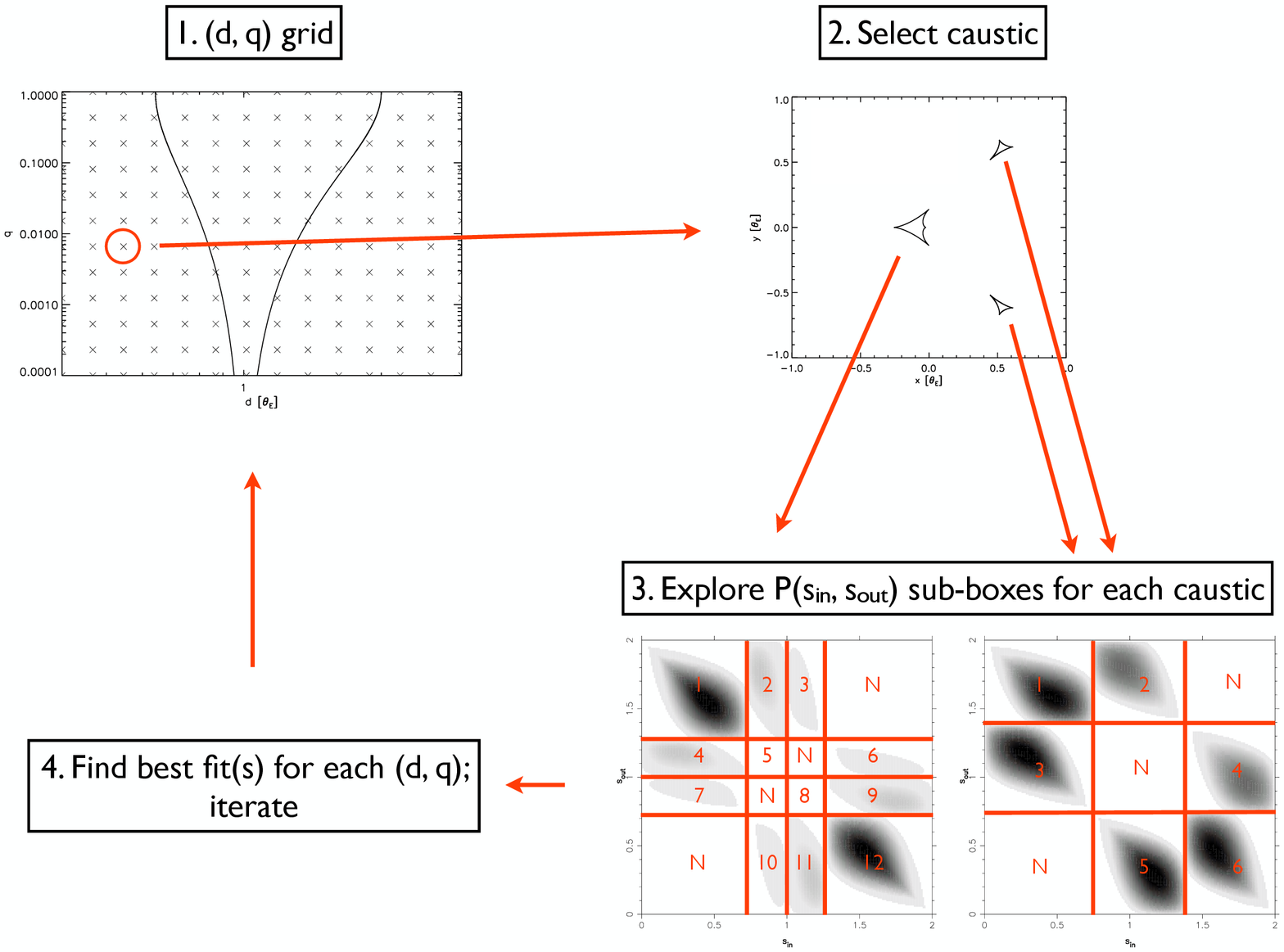}

  \caption{Algorithm flow for our modelling scheme. In the bottom right panel, 
``N" denotes an empty sub-box, while the other non-empty sub-boxes are 
enumerated. \label{fig:algorithm_flow}}

\end{figure*}

The flowchart in \Fig{fig:algorithm_flow}
summarises the main steps of our automated modelling scheme.
In summary, 
\begin{enumerate}

\item For each node in the $(d,q)$ grid, we 
construct the corresponding caustic curves.

\item For each caustic curve, we construct the $\priorp{\sin,\sout}$ prior map,
which divides into sub-boxes.

\item In each sub-box, we launch an MCMC run on the $\beta$ parameters
to find the best fit and map out the posterior 
$P(\beta|D,d,q,{\rm box})$. Chains are kept confined to each sub-box by forcing MCMC steps to 
remain within its boundaries.

\end{enumerate}

The results are then collected to construct 
the posterior probability $P(d,q|D)$, either by optimising the nuisance parameters or by integrating over the nuisance parameters in each sub-box, and then summing over the sub-boxes. Finally, we compute the corresponding ``Badness-of-Fit'' statistic
BoF$(d,q)=-2\,\ln{P(d,q|D)}$.

% =====================================================
% =====================================================

Our automated modelling scheme exploits the structure of the prior 
$\priorp{\sin,\sout}$.
For a caustic with $N_c$ cusps, the prior $\priorp{\sin,\sout}$ has 
$N_c\,(N_c-1)$ local maxima, one in each of the sub-boxes.
As the separation between these local maxima can be large, 
a single MCMC run, or whichever other parameter optimisation method is employed,
may find it difficult to jump from one sub-box to another, and thus 
may fail to find the best solution.
To avoid this, we launch an MCMC run in {\it each} of the sub-boxes.
Thus we start $N_c (N_c-1)$ chains, each confined to a particular sub-box,
to locate the best fit in each sub-box. We stop the chains using 
the convergence criterion of 
\cite{geweke92} once they are past a minimum number of iterations.

Our method thus divides the binary-lens parameter space not only into
a $(d,q)$ grid, but further into the required sub-boxes for 
each caustic for each $(d,q)$ pair.
In each sub-box we start the MCMC run at the maximum of $\priorp{\sin,\sout}$.
Instead of using $\chi^2$ as the sole criterion for acceptance or rejection 
of proposed MCMC steps, the ratio of the priors is also taken into account.

To incorporate non-uniform priors $\priorp{\theta}$ in the MCMC algorithm,
we simply modify the criterion for accepting a proposed step.
Rather than only the likelihood $P(D|\theta)$,
we consider the full posterior
$P(\theta|D)\propto\,P(D|\theta)\,\priorp{\theta}$.
A proposal to take a random step from $\theta$ to $\theta'$
is always accepted if $\theta'$ increases the posterior, 
and the acceptance probability when $\theta'$ diminishes 
the posterior is the ratio of posterior probabilities

 \begin{equation}\label{eq:acceptance}
	\frac{P(\theta'|D)}{P(\theta|D)}
 = \exp{\left(-\frac{1}{2}\Delta\chi^2\right)}\,
	\frac{\priorp{\theta'}}{\priorp{\theta}}
\ ,
\end{equation}

\noindent
where $\Delta\chi^2=\chi^2(\theta')-\chi^2(\theta)$ is the difference in $\chi^2$ 
across the proposed step.
For an MCMC run over the full parameter set $\theta$,
the relevant prior is
\begin{equation}
\label{eqn:priortheta}
	\priorp{\theta} = \priorp{d,q}\,
	\priorp{\beta|d,q}
\ .
\end{equation}
For MCMC runs over the nuisance parameters $\beta$, 
with fixed $d,q$, the relevant prior is
\begin{equation}
\label{eqn:priorbeta}
	\priorp{\beta|d,q} \propto \priorp{\sin,\sout}\,\phit(d,q,\alpha)
\ .
\end{equation}

\subsection{Implementation}

We implemented the algorithm by using a cluster of desktop computers, each one
running one of the MCMC chains to map out the
posterior $P(\beta|D,d,q,{\rm box})$ for a grid of $(d,q)$ values
and for all the corresponding sub-boxes.
The results are then collected to construct
the posterior probability map $P(d,q|D)$,
integrating over the nuisance parameters $\beta$, and
summing over the sub-boxes.

$P(d,q|D,{\rm box})$ is evaluated from each MCMC chain using the best-fit parameters $\hat{\beta}$:

\begin{equation}
\begin{array}{rl}
	P(d,q|D,{\rm box})
& \bs	\propto P(D|d,q,{\rm box},\hat{\beta})\,
	\priorp{\hat{\beta}|d,q,{\rm box}}\,
\ .
\end{array}
\end{equation}
Because each sub-box has its own MCMC chain, we must
weight the chain averages by the prior probability of each sub-box:
\begin{equation}
\priorp{{\rm box}|d,q} = \int\bs\int \priorp{\sin,\sout}\,
	\phit\, d\sin\, d\sout
\ ,
\end{equation}
where the integration limits cover the sub-box.
The weighted sum of chain averages 
then gives the posterior $(d,q)$ map,
\begin{equation}
P(d,q|D) =
	\displaystyle\sum\limits_{\rm box}
	\priorp{{\rm box}|d,q}\,
	P\left(d,q|D,{\rm box}\right)
\ .
\end{equation}
Normally one sub-box dominates the sum, but sometimes two or more
can contribute.

\section{Results}

\subsection{Badness-of-Fit Criteria}

We consider and compare results for four alternative
``Badness-of-Fit'' criteria,
corresponding to maximum likelihood (ML),
maximum a-posteriori (MAP), and the Bayesian Information Criterion (BIC), as well as a Bayes statistic that integrates  the posterior
probability over the nuisance parameters.
In each case the best-fit parameters $(d,q)$ minimise
a ``Badness-of-Fit'' statistic, BoF$(d,q)$,
and the corresponding posterior probability is
$P(d,q|D)\propto\exp{\left\{-\frac{1}{2}\,{\rm BoF}(d,q) \right\}}$.
Figs \ref{fig:chi2map}-\ref{fig:bayemap} display
the BoF$(d,q)$ maps obtained for the four cases:

\begin{eqnarray*}
 & \mathrm{\bf ML:} & \mathrm{BoF} = \chi^2(\hat{\beta})\\ 
 & \mathrm{\bf MAP:} & \mathrm{BoF} = \chi^2(\hat{\beta}) - 2\, \ln{ \prior(\hat{\beta}) }\\
 & \mathrm {\bf BIC:} & \mathrm{BoF} = \chi^2(\hat{\beta})
	- 2\, \ln{ \prior(\hat{\beta}) }
	+ \ln{ \left( N_{D} \right)}\,
		N_{\rm eff}\\		
& \mathrm{\bf Bayes:} & \mathrm{BoF} = \chi^2 - 2\, \ln({\prior\left({\beta}\right)\,d^m{\beta}})		
\ .
\end{eqnarray*}

Here the prior $\prior$ is from Eqn.~\ref{eqn:priortheta}
or Eqn.~\ref{eqn:priorbeta}, depending on the context.
We briefly elaborate on the three options below
before discussing the results.

\begin{itemize}

\item {\bf ML}:
 The {\it maximum likelihood} (ML) parameters
maximise the likelihood $L(d,q)=P(D|d,q)$, 
equivalent to minimising BoF$=-2\ln(L)=\chi^2$.
Thus we determine the best-fit value of $\chi^2$
for each $(d,q)$ pair, and let $P(d,q|D)\propto 
\exp{\left(-\chi^2/2\right)}$.
This approach emphasises the fit to the data
while disregarding priors on the parameters.

\item {\bf MAP}:
 The {\it maximum a-posteriori} (MAP) parameters
maximise the posterior probability density
$P(d,q|D)$, equivalent to minimising
BoF$(d,q)=-2\ln{P(d,q|D)}$.

\item {\bf BIC}: 
 The {\it Bayesian Information Criterion} (BIC),
applies an ``Occam'' penalty that gives
priority to ``simpler'' models that employ
fewer parameters to achieve their fit.
Each $(d,q)$ grid point is regarded as a competing model
with equal prior probability and $N_{\rm eff}$ effective parameters
that have been optimised.
The {\it Akaike Information Criterion} (AIC)
uses an Occam penalty $2\,N_{\rm eff}$ \citep{akaike74}, while the 
{\it Bayesian Information Criterion} (BIC) uses a
stronger penalty $\ln(N_D)\,N_{\rm eff}$, with $N_{D}$ the number of data points
\citep{schwarz78}. 
Our tests with fitting of polynomial models suggest that the BIC may 
be more reliable than the AIC for model selection.

We use the MCMC samples to estimate the ``effective number'' of nuisance 
parameters,
\begin{equation}
	N_{\rm eff} \approx \left<D(\theta)\right> - D(\left<\theta\right>)
\ ,
\end{equation}

\noindent
where $\left<x\right>$ denotes the expectation value of $x$ under the posterior, 
simply calculated by taking an unweighted average over the MCMC samples, and 
the `deviance' is
\begin{equation}
	D(\theta) \equiv \chi^2 - 2\,\ln{\prior}
\ ,
\end{equation}
as used to compute 
the acceptance probability of each step in the MCMC algorithm, as per 
\Eq{eq:acceptance}. Here $D(\left<\theta\right>)$ estimates the deviance at the 
minimum, while $\left<D(\theta)\right>$ measures the typical value, which 
should rise by 1 for each dimension of the parameter space explored by the MCMC
samples.
This definition of $N_{\rm eff}$ is designed to avoid double-counting when two 
parameters are highly correlated, and is found in the {\it Deviance Information 
Criterion} (DIC, \citealt{spiegelhalter02}, see also \citealt{ando07}).

\item {\bf Bayes}:
A fully Bayesian approach integrates the posterior probability density
over the $m$ nuisance parameters $\beta$, rather than just finding the maximum likelihood (ML) or maximum posterior probability density (MAP).
Thus if two models have the same MAP statistic, the one that achieves
that good fit over a wider range of parameters has a correspondingly higher
probability.

We can also write out the Bayes statistics as

\begin{equation}
\mathrm{BoF} = \chi^2(\hat{\beta}) - 2 \ln{ \prior( \hat{\beta} ) }
	-  \sum_{i=1}^m \ln{(2\,\pi\,\lambda_i( \hat{\beta} )) }\, ,
\end{equation}

\noindent
where the $2\,\pi$ factor here refers to the constant $\pi=3.141592...$ rather than the prior $\prior(\beta)$, and $\lambda_i( \hat{\beta} )$ are the $m$ eigenvalues of the parameter-parameter covariance matrix evaluated at $\hat{\beta}$,
their product being the $m$-dimensional parameter volume admitted
by the data around the best-fit value $\hat{\beta}$.

We approximate the integral over the $m$ nuisance parameters by
the method of steepest descents, 

\begin{eqnarray}
	\int e^{-\chi^2(\beta)/2}\,\prior\left(\beta\right)\,d^m\beta
\approx e^{-\chi^2(\hat{\beta})/2}\,\prior(\hat{\beta})\,d^m\beta \\
\approx e^{-\chi^2(\hat{\beta})/2}\,\prior(\hat{\beta})\, \prod_{i=1}^m \left(2\,\pi\lambda_i(\hat{\beta})\right)^{1/2}\, ,
\end{eqnarray}

\noindent
where the $2\,\pi$ factor here again refers to the constant rather than the prior.
This is just the MAP statistic multiplied by a parameter space volume.
We evaluate parameter space volume $d^m\beta$ as the square root
of the determinant of the parameter-parameter covariance matrix derived
from the MCMC chain.

\end{itemize}

\subsection{Fits to synthetic data}

\begin{figure*}
  \centering
  \includegraphics[width=8cm, angle=0]{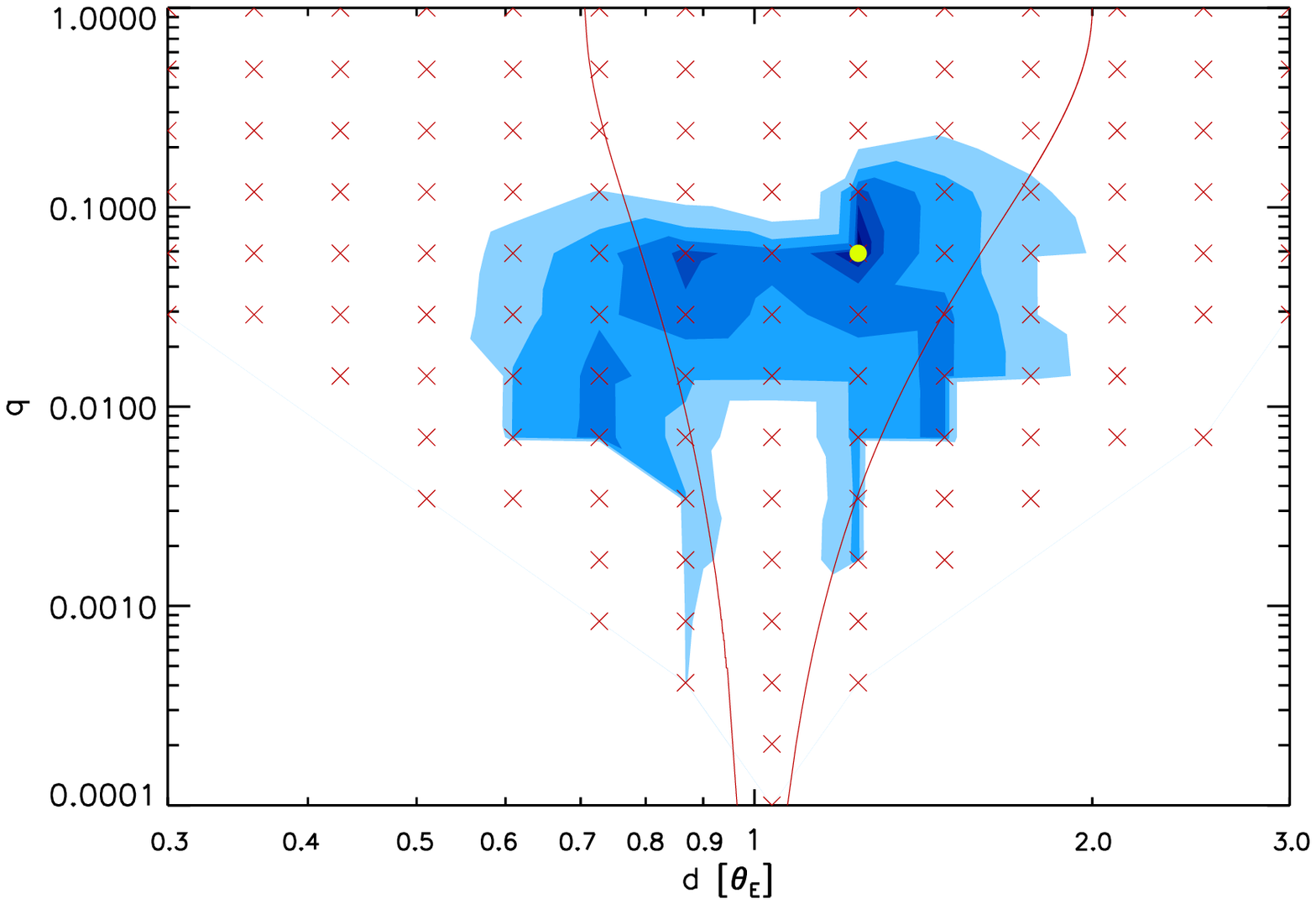}
  \includegraphics[width=8cm, angle=0]{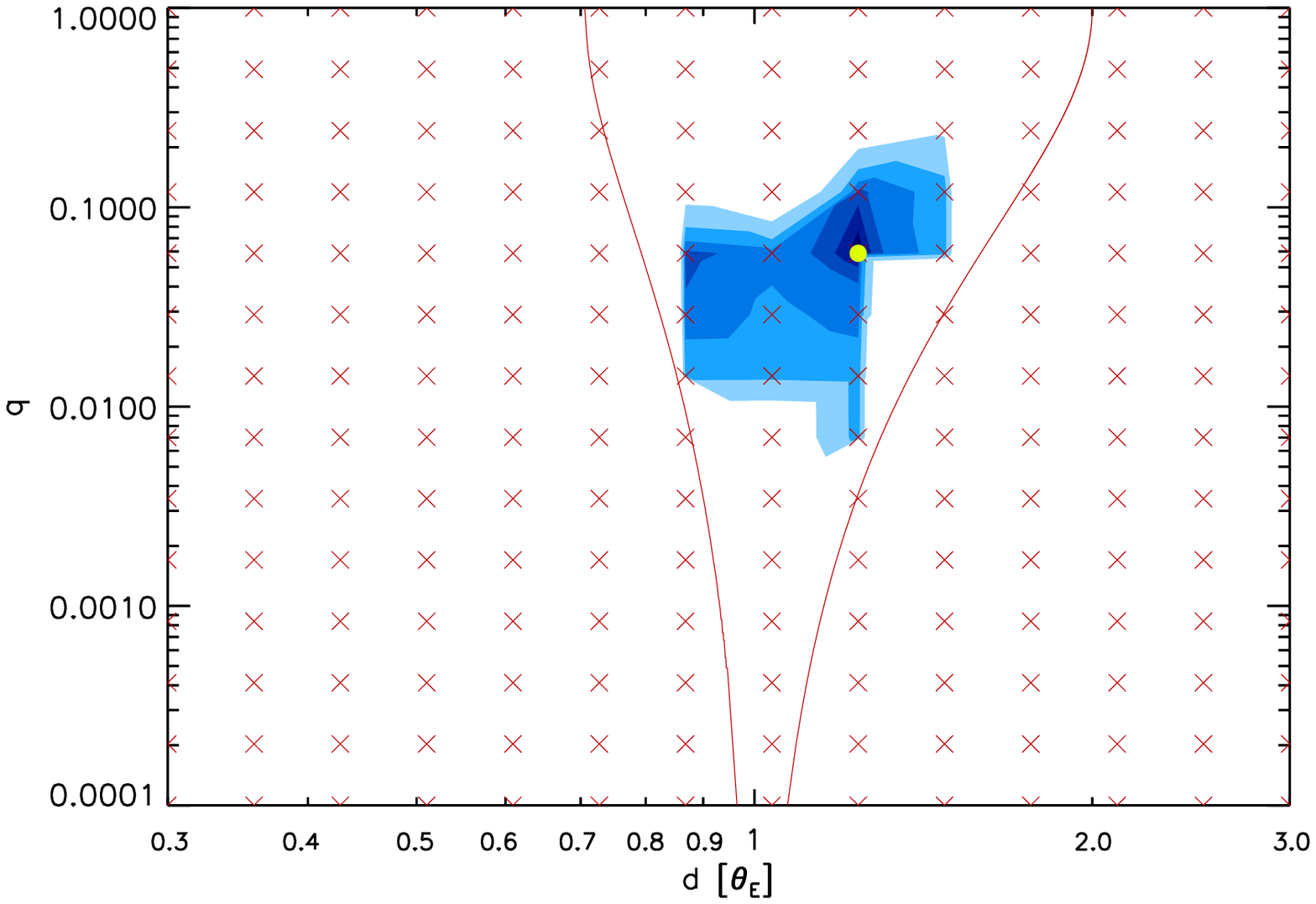}
  \includegraphics[width=6cm, angle=270]{fig/OB010888-model1.ps}
 \hspace{1cm} \includegraphics[width=6cm, angle=270]{fig/pmap-synth.ps}

\caption{MAP fit to the synthetic lightcurve data, using 
BoF$=\chi^2(\hat{\beta}) - 2\, \ln{ \prior(\hat{\beta}) }$. 
 \textit{Top}: 
 Posterior maps $P(d,q|D)$ for the source crossing a 
central (left) and secondary (right) caustic. 
Contour levels are at $\Delta\mathrm{BoF} = 2.3, 6.17, 
11.8, 20, 50, 100, 250$. The model with the lowest BoF is 
marked with a yellow filled circle and that of the true model with 
a filled blue triangle.
 \textit{Bottom}:
 The data and best-fit model lightcurve, 
with an inset showing the source trajectory crossing the caustic (left), and the 
location (yellow filled circle) of the best-fit model on the 
prior map $P(\sin, \sout)$,
along with MCMC samples (red circles) and local best-fit minima in each sub-box, 
indicated by yellow circles (right).
\label{fig:888_bof2map}}

\end{figure*}

For the synthetic event there is a single narrow, well-defined minimum. 
Table \ref{tab:888_parameters} summarises the parameters of the best-fit model 
found with a $14\times14$ ($d, q$) grid, evenly spaced in $\log{d}$ and 
$\log{q}$.
The posterior distribution $P(d,q|D)$ found
using the MAP option is plotted in \Fig{fig:888_bof2map}; 
the ML and BIC posterior maps are almost undistinguishable. 
The BoF minimum is so tightly defined that the choice of BoF statistic
has little effect on the best-fit parameters or the
shape of the posterior map.

The fit that is recovered it located at the grid point closest to the true model, as can be seen by comparing \Fig{fig:888_bof2map} to \Fig{fig:888_truelc}, and the best-fit parameters  given in \Tab{tab:888_parameters} to \Tab{tab:888_trueparameters}. The true parameters are not exactly recovered, as they do not match our grid points, but another modelling could be conducted without keeping $d$ and $q$ fixed, using the best models for each grid point as a starting point for new MCMC runs.

% =====================================================
% =====================================================

% =====================================================
\begin{table}
\begin{center}
  \begin{tabular}{ccc}
    \hline
    Parameter & & Units
\\ \hline  
   $d$ (grid) & $1.237$ & $-$
\\ $q$ (grid) & $0.059$	& $-$
\\ $g=\fb/\fs$ & 5.81 $\pm$ 0.09 & $-$
\\ $\chi^2$ & 202.9 & $-$
\\ \hline ``Standard"
\\ \hline
  $\tz$ & $5503.62 \pm 0.014$ & MHJD
\\ $\te$ & $37.52 \pm 0.04$ & days
\\ $\alpha$ & $1.692 \pm 0.005$ & rad
\\ $\uz$ & $0.056 \pm 0.003$ & $-$
\\ $\rhostar$ & $(2.44 \pm 0.26) \times 10^{-3}$ & $-$
\\ \hline ``Caustic"
\\ \hline
   $\tin$ & 5500.394 $\pm$ 0.009 & MHJD
\\ $\tout$ & 5507.342 $\pm$ 0.004 & MHJD
\\ $\sin$ & 1.273 $\pm$ 0.002 & $-$
\\ $\sout$ & 0.706 $\pm$ 0.002 & $-$
\\ $\dtcc$ & 0.092 $\pm$ 0.010 & days
\\ \hline \hline
  \end{tabular}
  \caption{Best-fit parameters (from $d, q$ grid exploration) for the synthetic 
   event. \label{tab:888_parameters}}
  \end{center}
\end{table}
% =====================================================

\subsection{Fits to OGLE-2007-BLG-472 data}

\begin{figure*}
  \centering
  \includegraphics[width=8cm, angle=0]{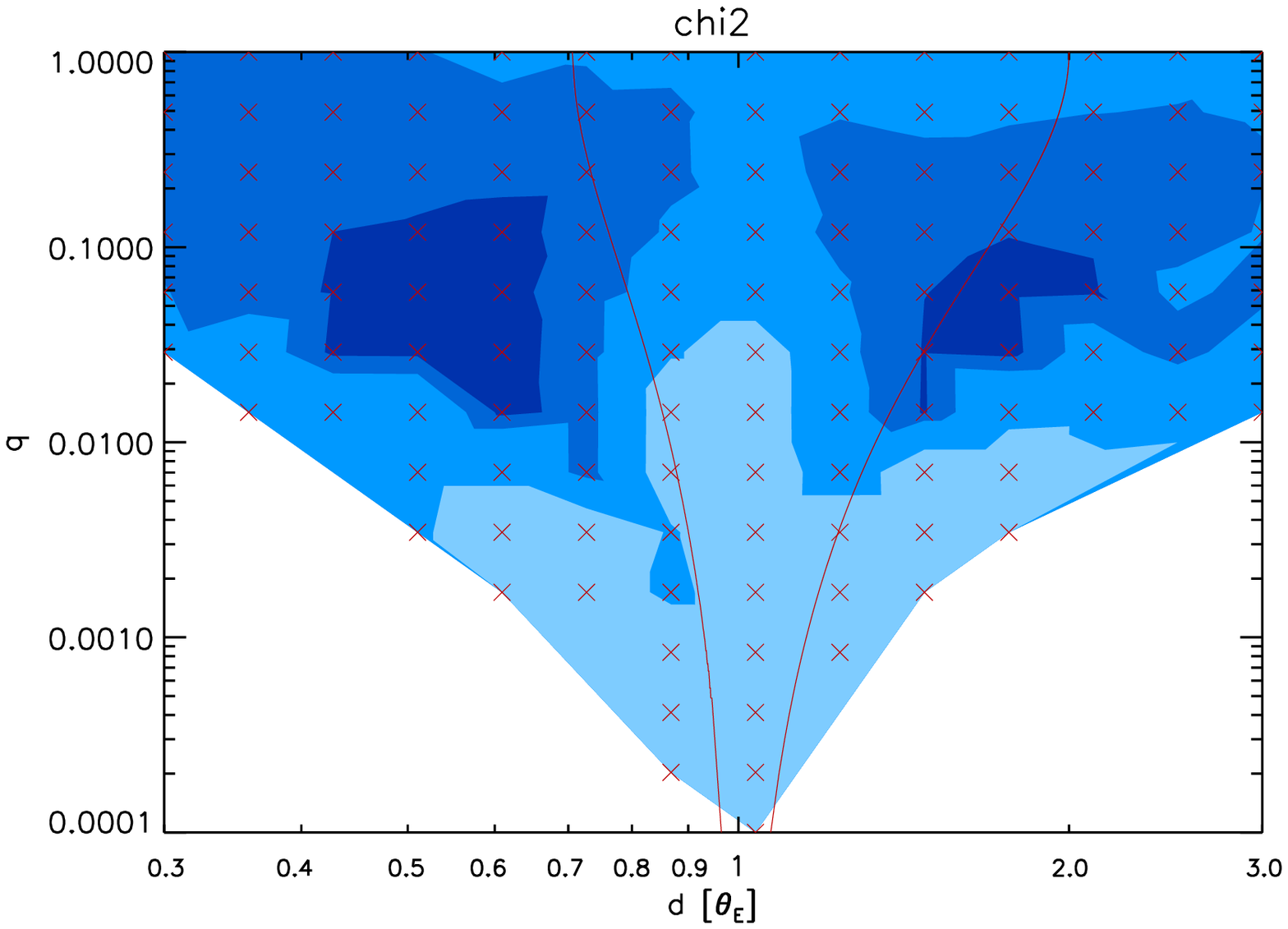}
  \includegraphics[width=8cm, angle=0]{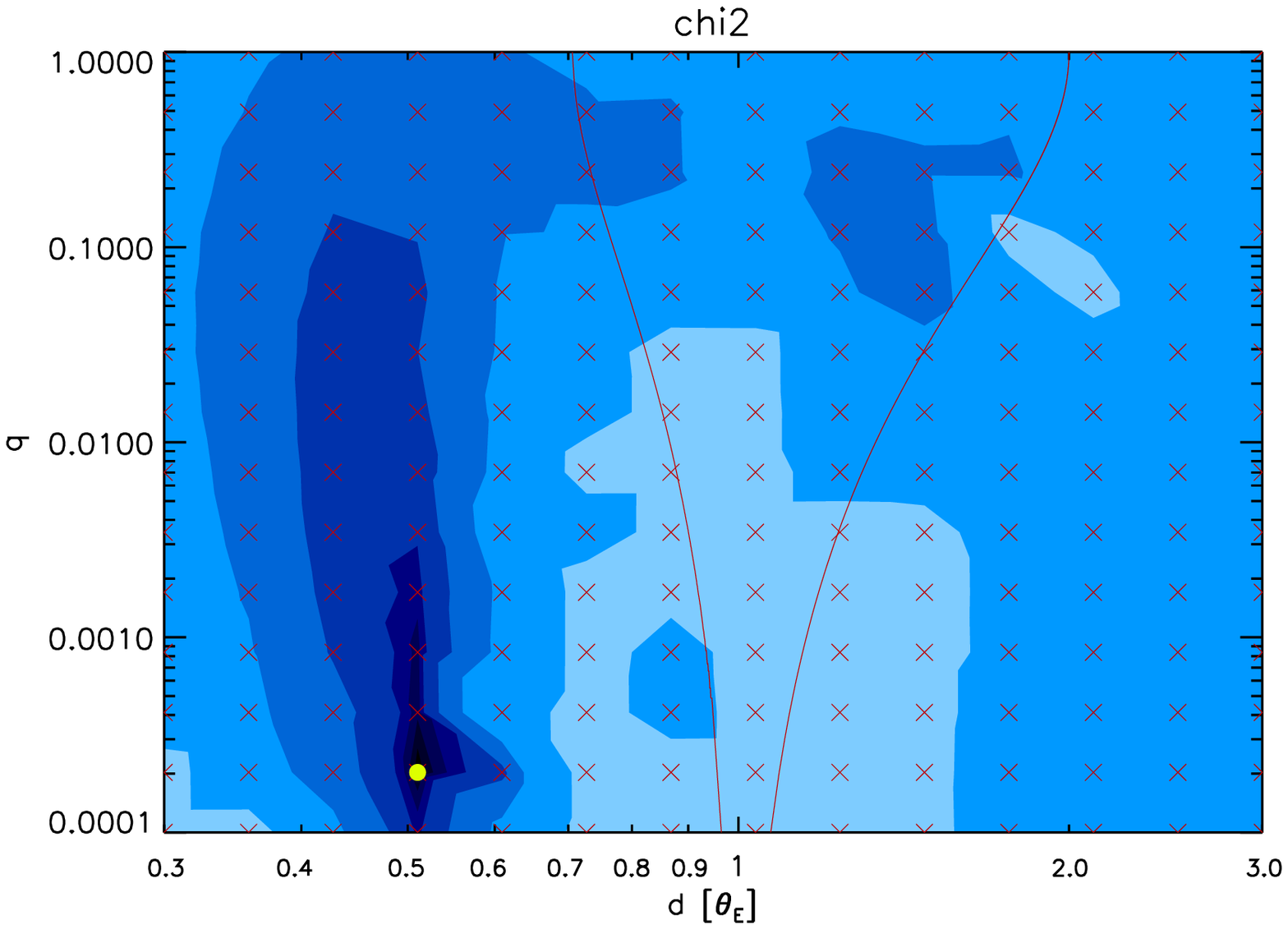}
  \includegraphics[width=6cm, angle=270]{fig/OB070472-model1.ps}
 \hspace{1cm} \includegraphics[width=6cm, angle=270]{fig/OB070472-model1-pmap-chains.ps}

  \caption{ML fit to the OGLE-2007-BLG-472 data,
using BoF$=\chi^2$.
 \textit{Top}: Posterior maps $P(d,q|D)$ for the source crossing
the central caustic (\textit{left}) and the secondary caustic (\textit{right}). 
A filled yellow circle marks the location of the model with the lowest $\chi^2$,
and the contour levels are at 
$\Delta\chi^2 = 2.3, 6.17, 11.8, 20, 50, 100, 250$.
\textit{Bottom}: 
The data and best-fit model lightcurve (\textit{left}), with an inset 
showing the source trajectory and the caustic, and 
the prior map $P(\sin, \sout)$ (\textit{right}),
along with the MCMC samples (red circles) and local 
best-fit minima (yellow circles) in each sub-box.
The best-fit model is marked by a blue filled circle.
\label{fig:chi2map}}

\end{figure*}

\begin{figure*}
  \centering
  \includegraphics[width=8cm, angle=0]{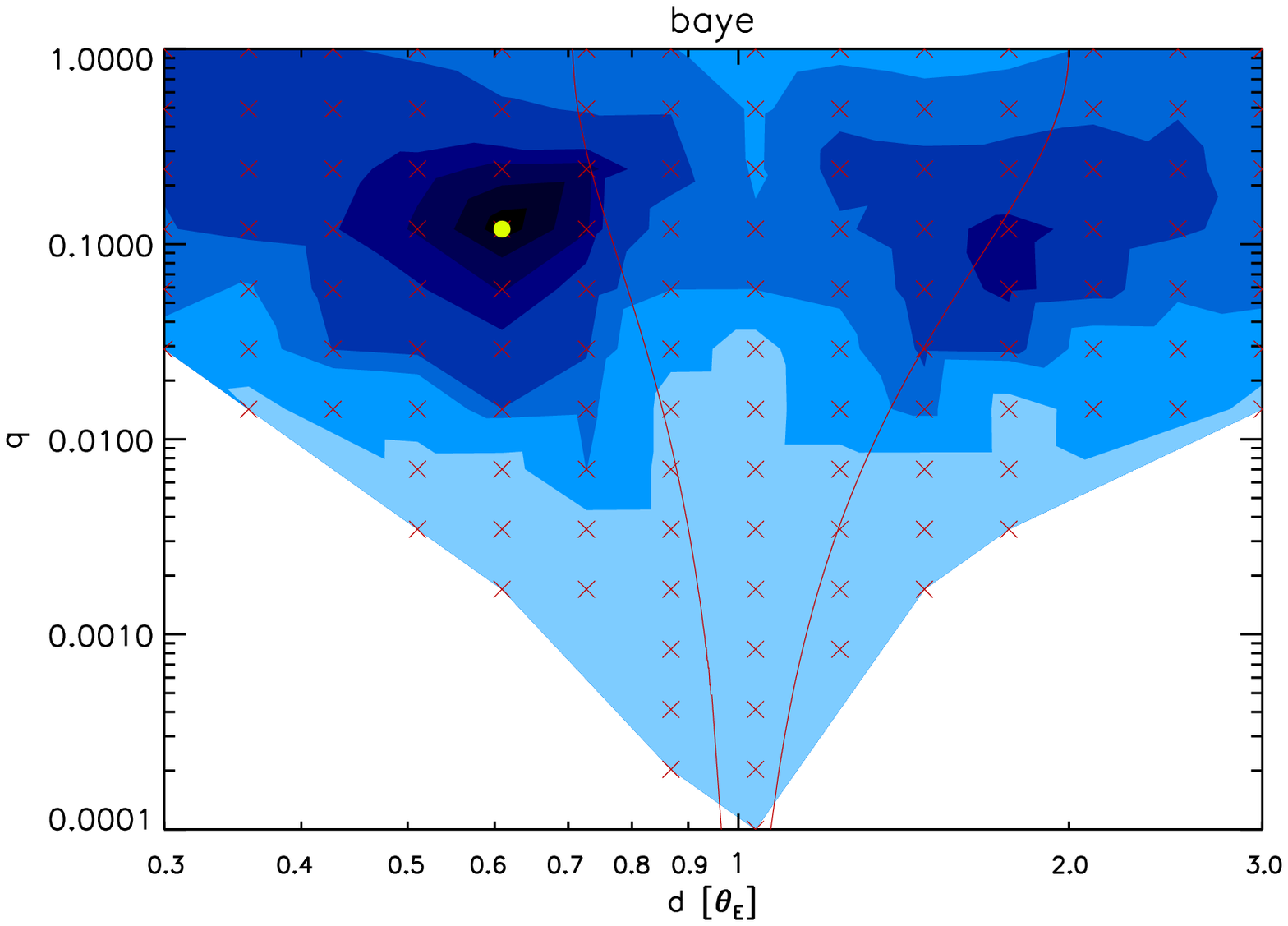}
  \includegraphics[width=8cm, angle=0]{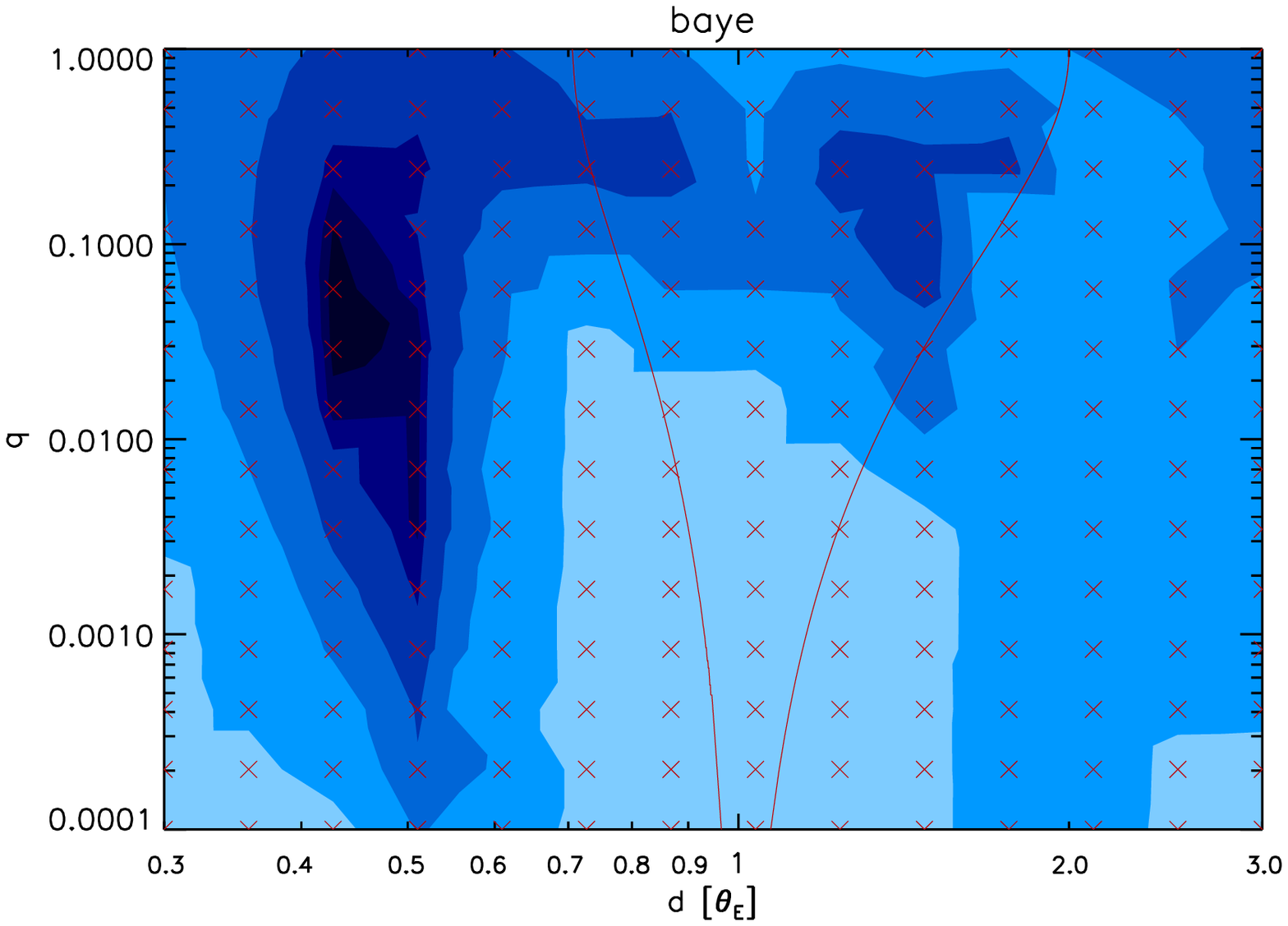}
  \includegraphics[width=6cm, angle=270]{fig/OB070472-model2.ps}
 \hspace{1cm} \includegraphics[width=6cm, angle=270]{fig/OB070472-model2-pmap-chains.ps}
    
  \caption{Same as \Fig{fig:chi2map} but for a fully Bayesian fit using 
 BoF$= \chi^2 - 2\, \ln({\prior\left({\beta}\right)\,d^m{\beta}})$, thus augmenting the likelihood
with suitable Bayesian priors on the parameters and taking into account the effective number of parameters, as detailed in the text. Corresponding figures for a MAP and BIC fit are shown on \Fig{fig:bayemap} for comparison.
 \label{fig:bicmap}}

\end{figure*}

\begin{figure*}
  \centering
  \includegraphics[width=8cm, angle=0]{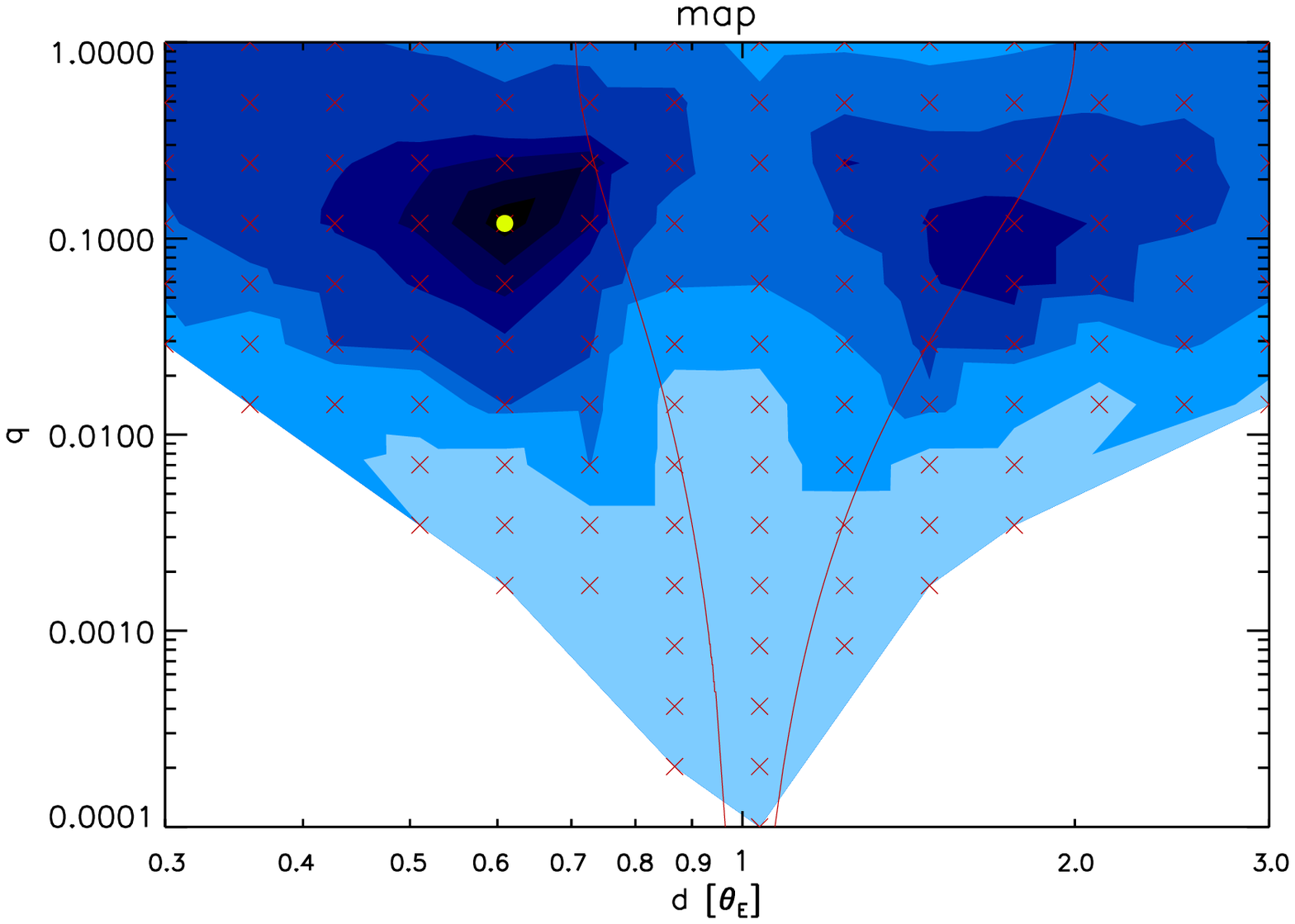}
  \includegraphics[width=8cm, angle=0]{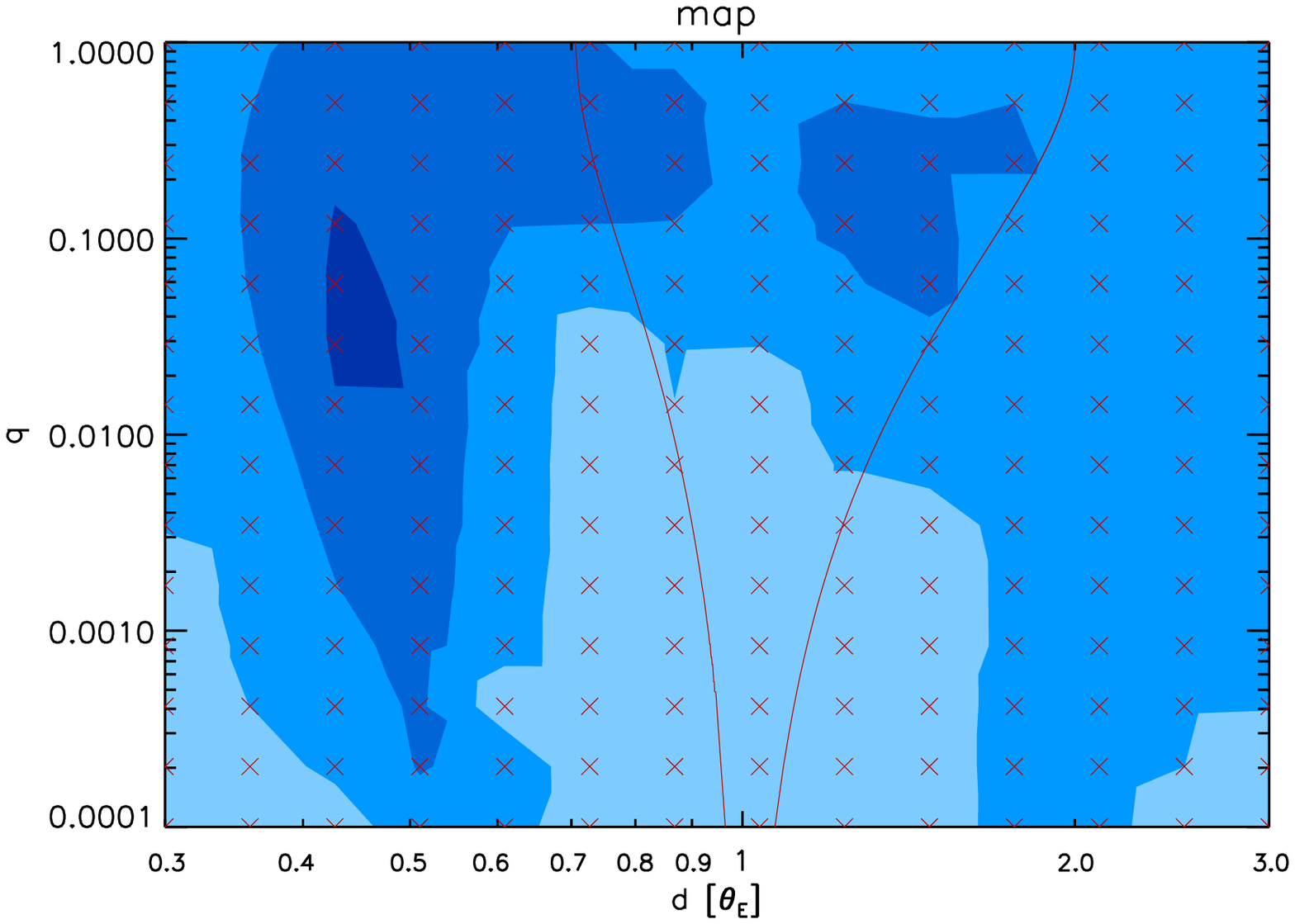}
  \includegraphics[width=8cm, angle=0]{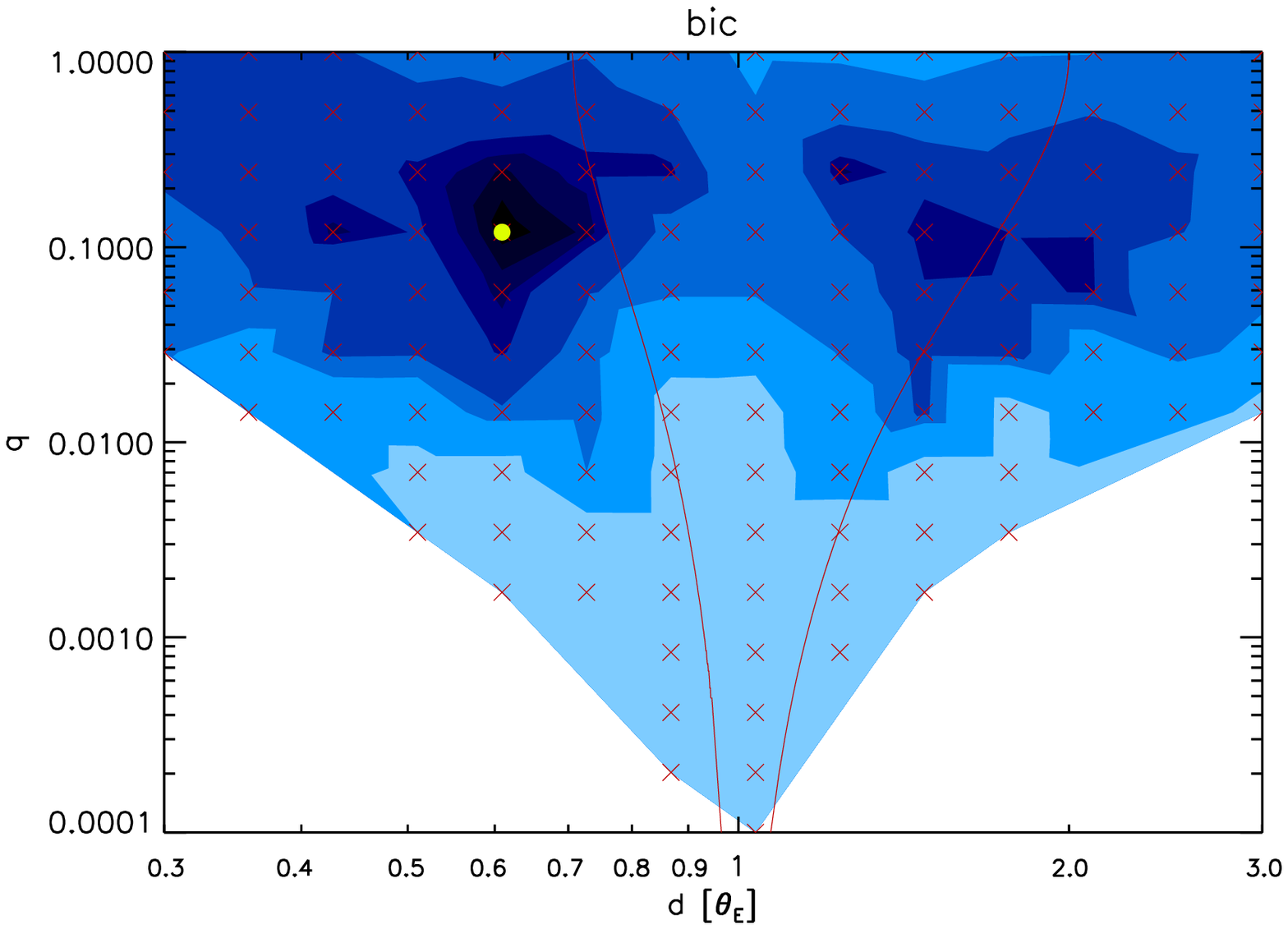}
  \includegraphics[width=8cm, angle=0]{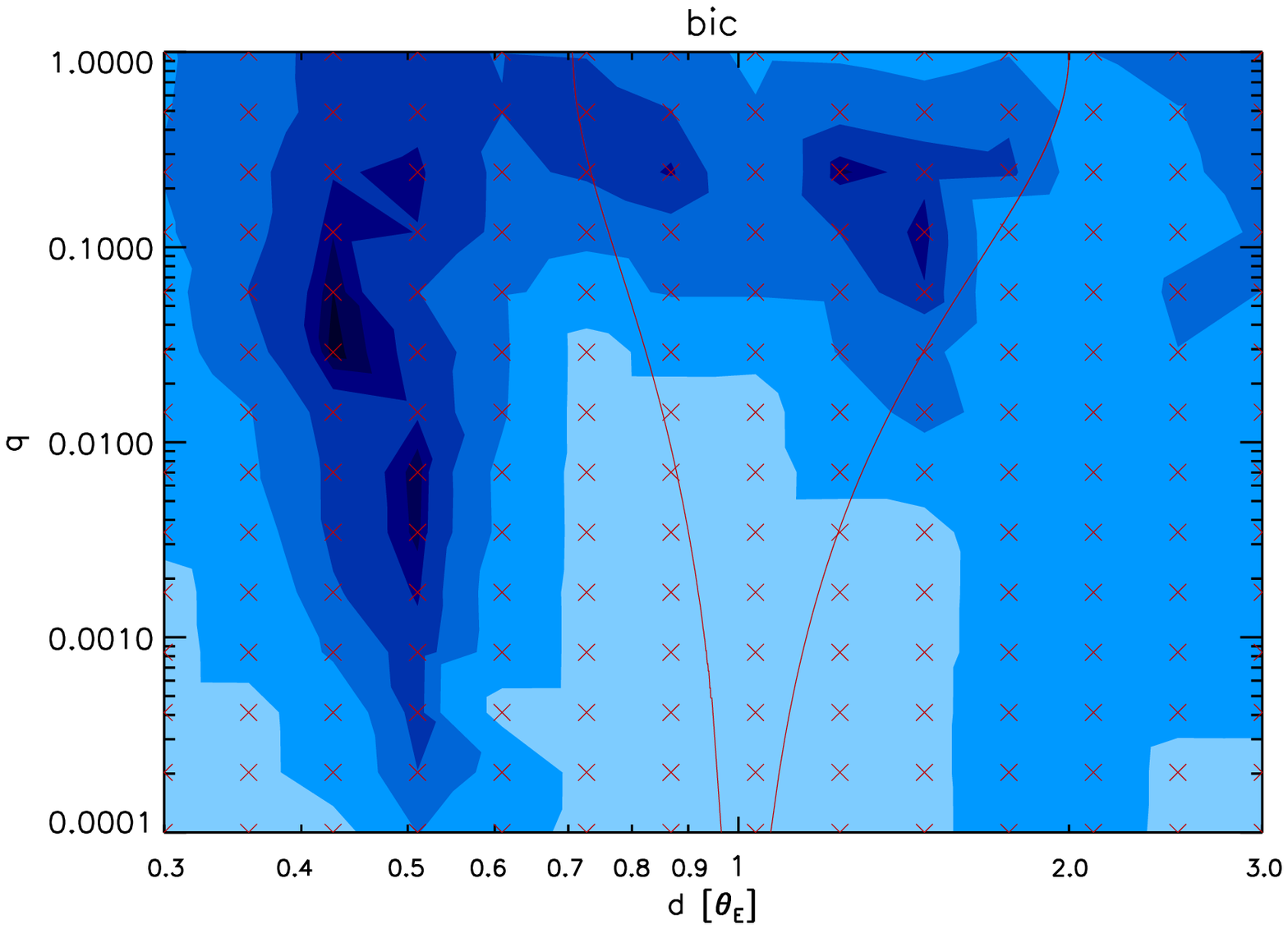}
    
  \caption{Posterior maps using the MAP and BIC statistics for comparison with the ``Bayes" map (Fig. \ref{fig:bicmap}); maps for central caustic crossings are shown at the left, with secondary caustic crossings shown on the right. \label{fig:bayemap}}

\end{figure*}

Our fits to the OGLE-2007-BLG-472 data are presented in
Figs~\ref{fig:chi2map}-\ref{fig:bayemap}.
The contour levels are set at $\Delta$BoF=2.3, 6.17, 11.8, 20, 50, 100, 250 and 500,
relative to the global minimum, the first 3 thus corresponding to 1, 2, and 
3-$\sigma$ confidence regions if the posterior is well
approximated by a 2-parameter Gaussian.
The best-fit values and uncertainties of additional
parameters are summarised in Table~\ref{tab:par_472}.

Fig.~\ref{fig:chi2map} exhibits an extended region of 
low $\chi^2$ around the minimum at $d=0.51$ and $q=2\times10^{-4}$.
The width in $d$ is unresolved by the rather coarse $(d,q)$ grid,
and the extension in $\log{q}$ is around 1~dex.
The best-fit model has the source crossing a 
very small planetary caustic, requiring a very long event timescale, 
$\te\sim2000$~d, to match the observed crossings at $\tin$ and $\tout$ separated 
by 3~d. Thus, as was also found in \cite{kains09}, the lowest-$\chi^2$ model for 
this event is not very well constrained, and has an implausibly long $\te$.
There are no significantly different competing local minima with $\Delta\chi^2<20$; the first competitive model for which the configuration (source trajectory and location of the caustic crossings) is significantly different has $\Delta\chi^2 \sim 22$.

Changing the BoF statistic has a significant effect on the 
posterior map: the penalties introduced by the prior move the 
best-fit model ``up" in $q$, towards models with smaller $\te$ and configurations where the source crosses a central, rather than planetary, caustic.
The MAP, BIC and Bayes fits (Fig.~\ref{fig:bicmap} - \ref{fig:bayemap}) all favour a model with $q\sim0.12$, $d\sim 0.61$.
 
The $\chi^2$ increases by $\Delta\chi^2=43$ relative to the global $\chi^2$ 
minimum, but the priors compensate since $\te$ and $\rho_*$ both move toward more
plausible values, and the larger caustic is easier to hit. \Fig{fig:priorcontour} shows the location of the lowest-$\chi^2$ and best Bayesian models with respect to the $\prior(\rho_*, \te)$ contour. The $\chi^2$ model is in the wings of the prior distribution, whereas the best-BIC model is near its peak, meaning that the $\chi^2$ model is more strongly penalised by the prior.

\begin{figure*}
  \centering
  \includegraphics[width=8cm, angle=0]{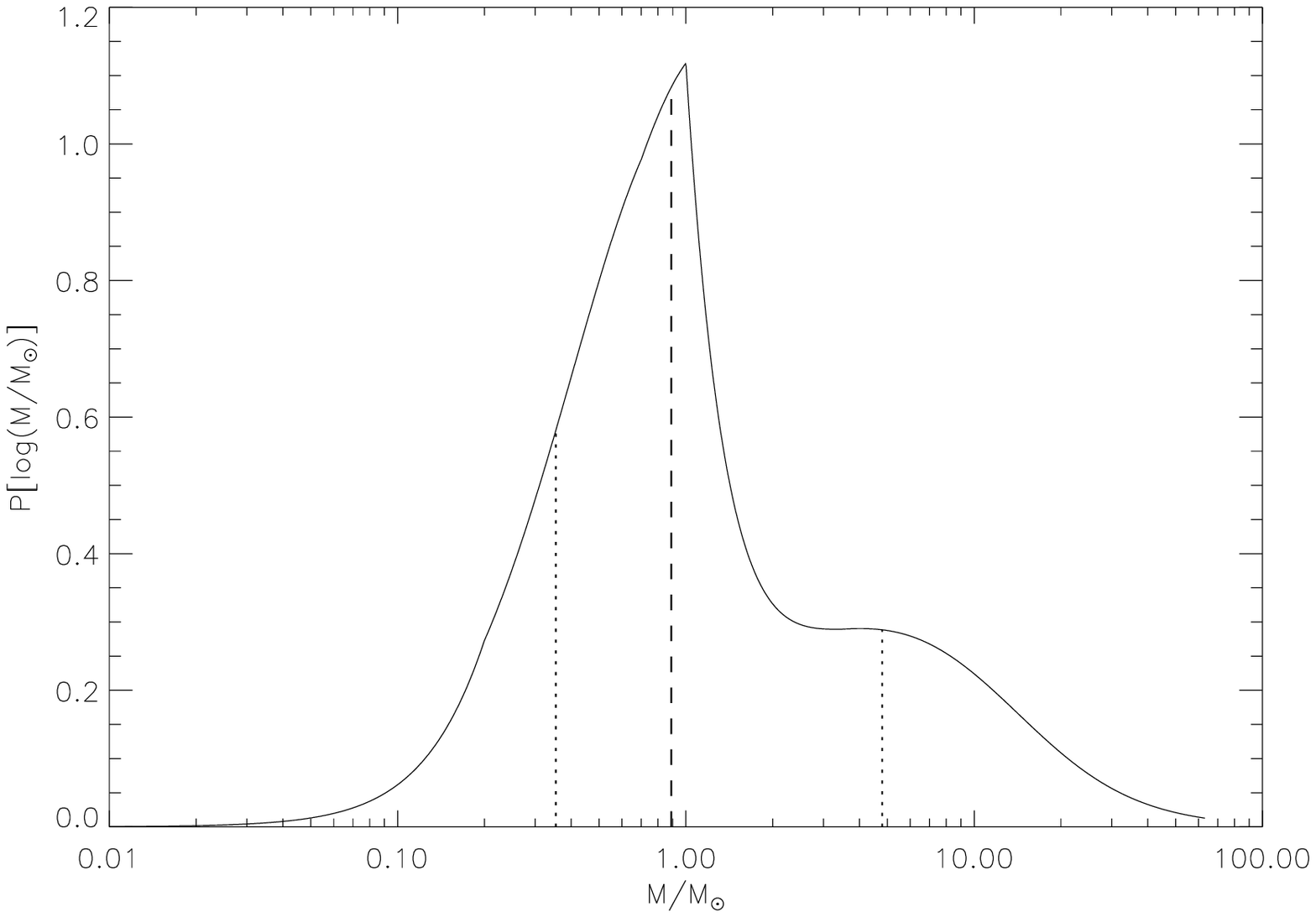}
  \includegraphics[width=8cm, angle=0]{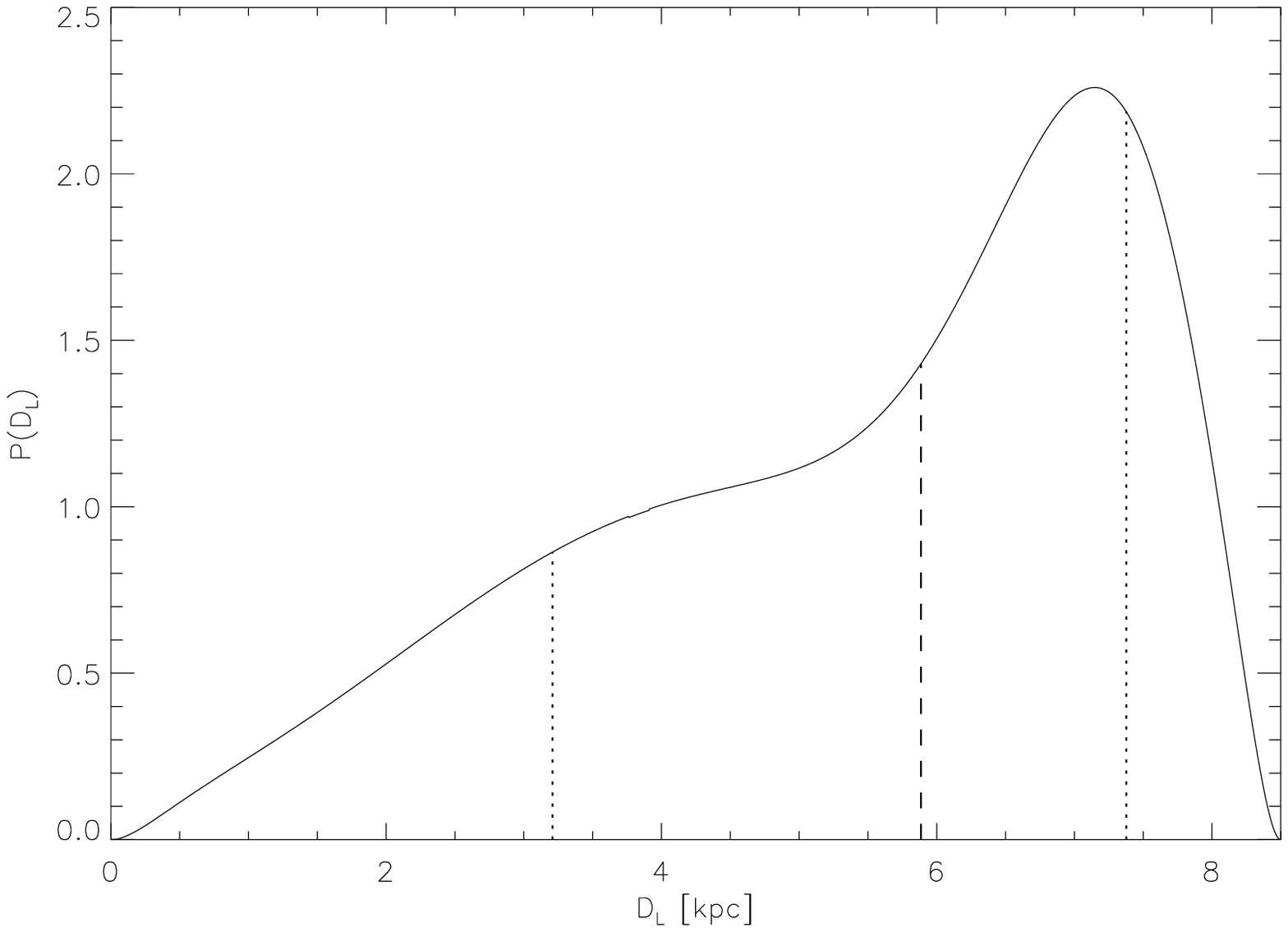}

  \caption{Probability density distributions for the lens mass (left) and distance 
(right) using the best-fit parameters found by minimising the BIC, 
computed using the 
algorithm of Dominik~(2006). The distance distribution assumes a source 
distance 
$D_{\rm S}=8.5$kpc. Dashed lines and dotted lines respectively denote the median 
value and the limits of the 68.3\% confidence interval. 
\label{fig:lensprop}}

\end{figure*}

We used the best-fit Bayesian parameters (Table~\ref{tab:par_472})
and the algorithm of \cite{dominik06}
to derive probability distributions for the lens mass and distance
as shown in \Fig{fig:lensprop}.
With no parallax signal detected in this event, 
we used only the constraint from $\te$ and $\rho_*$.
We find lens component masses of 
$0.78^{+3.43}_{-0.47}\,\msun$ and $0.11^{+0.47}_{-0.06}\,\msun$ at a distance of 
$5.88^{+1.49}_{-2.68}$ kpc. 

The best Bayesian model has a large blend/source flux ratio $F_B/F_S\sim200$.
There is no obvious star blended with the lens on the images, 
and the blending could plausibly come from the binary-star lens system,
or from a third body.

% =====================================================
% =====================================================

% =====================================================
% =====================================================

% =====================================================
% =====================================================

% =====================================================
\begin{table*}
\begin{center}
  \begin{tabular}{cccccc}
    \hline Parameter
	& ML (Fig. \ref{fig:chi2map})
	& MAP/BIC/Bayes (Fig. \ref{fig:bicmap})
	& Units
\\ \hline ML ($\chi^2$)
	& 915
	& 958
	& $-$
\\ MAP
	& 1017
	& 968
	& $-$
\\ BIC
	& 1046
	& 986
	& $-$
\\ Bayes
	& 1061
	& 1003
	& $-$ 
\\ \hline	
 $d$ (grid)
	& 0.51
	& 0.61
	& $-$
	
\\ $q$ (grid)
	& $2.03\times10^{-4}$
	& $0.119$
	& $-$
\\ $g=\fb/\fs$
	& 7.59 $\pm$ 0.08
	& $114.59 \pm 9.62$
	& $-$	
\\ (OGLE)
\\ $N_{\rm eff}$
	& 4.20
	& 4.97
	& $-$

\\ \hline Standard
\\ \hline $\tz$
	& $7121.28 \pm 113.61$
	& $4332.41 \pm 0.25$
	& MHJD
\\ $\te$
	&  $1939.35 \pm 80.92$
	& $73.37 \pm 5.45$
	& day
\\ $\alpha$
	& $3.134 \pm 0.044$
	& $3.050 \pm 0.020$
	& rad
\\ $\uz$
	& $-0.181 \pm 0.029$
	& $-0.052 \pm 0.003$
	& $-$
\\ $\rhostar$
	& $(3.09 \pm 0.37) \times 10^{-5}$        
	& $(5.66 \pm 0.51) \times 10^{-4}$        
	& $-$
\\ \hline ``Caustic"
\\ \hline $\tin$
	& 31.379 $\pm$ 0.012
	& 31.325 $\pm$ 0.016
	& MHJD-4300
\\ $\tout$
	& 34.078 $\pm$ 0.002
	& 34.077 $\pm$ 0.002
	& MHJD-4300
\\ $\sin$
	& 1.785 $\pm$ 0.012
	& 0.807 $\pm$ 0.005
	& $-$
\\ $\sout$
	& 1.011 $\pm$ 0.013
	& 0.423 $\pm$ 0.011
	& $-$
\\ $\dtcc$
	& 0.073 $\pm$ 0.003
	& 0.072 $\pm$ 0.004
	& day
\\ \hline \hline $I_s$
	& 17.95
	& 20.77
	& mag
\\ $I_b$
	& 15.74
	& 15.62
	& mag
\\ $\theta_*$
	& 1.15
	& 0.53				
	& $\mu$as
\\
\hline
  \end{tabular}

  \caption{Best-fit standard (top) and ``caustic" (middle) parameters for 
OGLE-2007-BLG-472, as well as source properties and blend magnitude (bottom), 
for the ML and BIC statistics. The values of all 4 BoF statistics are given for both models for informative purposes.
 $d$ and $q$ are fixed since these are grid models. Source 
angular radii are computed using the same colour as in the previous paper on this 
event (Kains et~al. 2009). \label{tab:par_472}}

  \end{center}
\end{table*}
% =====================================================

% =====================================================
% =====================================================

\section{Discussion and conclusions}\label{sec:conclusion}

The modelling results for the two datasets presented here indicate that our 
algorithm is successful in locating minima throughout the parameter space, and 
the subdivision of the prior maps ensures that all possible source trajectories 
through the caustics are explored. Furthermore, the use of Bayesian priors allows 
us to incorporate information on the event timescale distribution, as well 
geometrical information on the concavity of caustics.

Although the sampling rate for our synthetic lightcurve data is not 
particularly high 
compared to what can now be achieved by survey and follow-up teams,
our algorithm located a well-defined minimum near the true minimum, 
with a grid search of the $(d,q)$ parameter space and MCMC runs
to sample the posterior probability in the region around each
local minimum.

In our re-analysis of the OGLE-2007-BLG-472 data,
we improve upon the posterior map calculated in \cite{kains09} for 
OGLE-2007-BLG-472 because we now use an MCMC 
run for each prior sub-box separately rather than just a single one per 
$(d,q)$ grid point. 
We find that changing the badness-of-fit statistic leads to important 
changes in the posterior $P(d,q|D)$ maps.
In particular, the model with lowest $\chi^2$ has a planetary mass ratio
and an implausibly long $\te\sim2000$~d. 
Adding priors dramatically shifts the location of the best-fit model, lowering the timescale to $\te\sim70$~d.
Using a Bayesian approach to penalise models with improbable parameters 
leads to best-fit parameters corresponding 
a binary star lens with $0.78$ and 0.12~$\msun$ components
at a distance of $\sim5.9$~kpc, and a more typical
event timescale $\te\sim 70$~d.
The only remarkable parameter is a rather high blending fraction,
which could arise from either the lens itself or a closely blended third star.
The new model is very different from that found by \cite{kains09}, which 
characterised the lens as a binary star with 
components masses of $1.50$ and $0.12 \msun$ at a distance of 1 kpc.

The development of automated algorithms for real-time modelling such as that 
presented here allows observers to receive feedback on ongoing anomalous 
microlensing events, and ensure that important features predicted by real-time 
modelling are not missed. This makes it much easier to assess the nature of the 
lensing system more rapidly and allocate observing time to targets more 
effectively. When observational coverage is not complete, or when the  $\chi^2$ alone is not sufficient as a 
criterion for badness-of-fit, statistics like the 
those we use in this paper could help to assess reliably alternative models. Furthermore, provided that the chosen priors are appropriate, comparing the resulting posterior maps of using different statistics allows for a useful test of a given model's robustness.

\section*{Acknowledgments}

NK is supported by an ESO Fellowship. The research leading to these results has 
received funding from the European Community's Seventh Framework Programme 
(/FP7/2007-2013/) under grant agreement No 229517. 
KH and MH are supported by The Qatar Foundation QNRF grant NPRP-09-476-1-87.

\bibliographystyle{mn2e}
\bibliography{../thesisbib}
\bsp

\label{lastpage}

\end{document}